\documentclass{article}
\usepackage{amsmath}
\usepackage{amssymb}
\usepackage{amsfonts}
\usepackage{bm}
\usepackage[dvips,xdvi]{graphicx}
\usepackage[compress,numbers,super]{natbib}

\begin{document}



\title{\bf Plateau Moduli of Several Single-Chain \\
Slip-Link and Slip-Spring Models}
\author{Takashi Uneyama\textsuperscript{*} and Yuichi Masubuchi \\
\\
Center for Computational Science, Graduate School of
Engineering, \\
Nagoya University, \\
Furo-cho, Chikusa, Nagoya 464-8603, Japan \\
e-mail: uneyama@mp.pse.nagoya-u.ac.jp}
\date{}

\maketitle




\begin{abstract}
We calculate the plateau moduli of several single-chain slip-link and
slip-spring models for entangled polymers.
In these models, the entanglement
effects are phenomenologically modeled by introducing topological constraints
such as slip-links and slip-springs.
The average number of segments between two neighboring slip-links
or slip-springs, $N_{0}$, is an input parameter in these models.
To analyze experimental data, 
the characteristic number of segments in entangled polymers $N_{e}$
estimated from the plateau modulus is used instead.
Both $N_{0}$ and $N_{e}$ characterize the topological constraints in
entangled polymers, and 
naively $N_{0}$ is considered to be the same as $N_{e}$.
However, earlier studies showed that $N_{0}$ and $N_{e}$ (or the plateau modulus)
should be considered as independent parameters.
In this work, we show that due to the
fluctuations at the short time scale, $N_{e}$ deviates from $N_{0}$.
This means that the relation between $N_{0}$ and the plateau modulus 
is not simple as naively expected.
The plateau modulus (or $N_{e}$) depends on the
subchain-scale details of the employed model, as well as the
average number of segments $N_{0}$.
This is due to the fact that the subchain-scale fluctuation 
mechanisms depend on the model rather strongly.
We theoretically calculate the plateau moduli for
several single-chain slip-link and slip-spring models.
Our results explicitly show that the relation between $N_{0}$ and $N_{e}$
is model-dependent.
We compare theoretical results with various simulation data in the literature,
and show that our theoretical expressions reasonably explain the simulation results.
\end{abstract}


\section{Introduction}

Polymer melts and solutions with sufficiently large molecular weights
exhibit characteristic relaxation behaviors due to the ``entanglements.''
The entanglements originate from the non-crossability between polymer chains.\cite{Doi-Edwards-book}
Polymer chains cannot cross each other, and thus the motion of a polymer chain
is strongly constrained compared with freely crossable phantom chains.
As a result, we observe various characteristic
dynamical behaviors.
For example, the longest relaxation time becomes very long
and the relaxation modulus $G(t)$ of a well
entangled polymer becomes approximately constant as
$G(t) \approx G_{N}^{(0)}$ (with $G_{N}^{(0)}$ being a constant)
in a relatively long time range before the system fully relaxes.
The plateau modulus $G_{N}^{(0)}$ is an important parameter which characterizes
the entanglement effect\cite{Fetters-Lohse-Colby-2007}. The emergence of the plateau modulus can be
phenomenologically interpreted as the existence of a transient rubber-like network in
an entangled polymer system.
Experimentally, we can directly measure the plateau modulus $G_{N}^{(0)}$
if the molecular weight is sufficiently large and polymers are well-entangled.
Even if the molecular weight
is not sufficiently large, several methods have been developed to estimate
the plateau modulus from the linear viscoelasticity data\cite{Liu-He-vanRuymbeke-Keunings-Bailly-2006}.
Therefore, from the experimental view point,
we can say that the plateau modulus of a specific polymer can be determined straightforwardly.
If we assume that an entangled polymer simply behaves as an ideal rubber network,
we can estimate the average molecular weight between (virtual) cross-links.
This molecular weight can be interpreted as the molecular weight between entanglements 
(the molecular weight between two sequential transient
linking points) $M_{e}$.

However, the situation is not that clear in mesoscopic coarse-grained models.
In most of mesoscopic coarse-grained simulation models for entangled
polymers, the entanglements are phenomenologically implemented as some
transient objects (such as tubes and slip-links).\cite{Doi-Edwards-book}
In such models, the molecular weight between entanglements is
interpreted as an input parameter. 
We express the molecular weight between two entanglement points in
coarse-grained models as $M_{0}$.
Naively one
may expect that $M_{e}$ from the plateau modulus
is just the same as the molecular weight between two entanglement points
$M_{0}$ ($M_{e} = M_{0}$).
However, this is not true in general.
We should emphasize that $M_{0}$ cannot be experimentally
observed (unlike $M_{e}$).
The relation between $M_{0}$ and $G_{N}^{(0)}$ depends on the details of the employed model.
Although the the relation between
$M_{e}$ and $G_{N}^{(0)}$ has been
studied in detail for some specific
models\cite{Steenbakkers-Tzoumanekas-Li-Liu-Kroger-Schieber-2014}, this relation
is not clear for most of coarse-grained models.
As a result, to fit simulation data of mesoscopic coarse-grained models to
specific experimental data, sometimes we are required to use both $M_{0}$ and $G_{N}^{(0)}$
as independent fitting parameters\cite{Masubuchi-Ianniruberto-Greco-Marrucci-2008,Masubuchi-Uneyama-2018}.
This procedure works well for practical purposes, but it seems not to be
fully consistent with a naive and
intuitive interpretation of $M_{0}$.

We should carefully consider the relation between $M_{0}$ and $G_{N}^{(0)}$
in mesoscopic coarse-grained models.
For example, in the tube model, the value of $G_{N}^{(0)}$ is known to be 
reduced if we take into account the longitudinal motion of a chain along the tube \cite{Doi-Edwards-book}.
If we interpret an entangled polymer system as a network,
the functionality of this network also affects the plateau
modulus, in the same way as the phantom network\cite{Everaers-1999,Masubuchi-Ianniruberto-Greco-Marrucci-2003}.
In this work, we theoretically investigate the plateau modulus for
various slip-link and slip-spring type models in detail.
Although the plateau modulus apparently seems to be well-defined, from the theoretical
view point, how to determine the plateau modulus is not fully clear.
In this work, we define the plateau modulus $G_{N}^{(0)}$ from the
relaxation modulus $G(t)$ as:
\begin{equation}
 \label{plateau_modulus_and_relaxation_modulus}
 G_{N}^{(0)} \equiv G(t \approx \tau_{e}) .
\end{equation}
Here, we should use $G(t)$ for a well entangled system of which
molecular weight is sufficiently large.
This definition is almost the same as the ``entanglement modulus $G_{e}$''
proposed by Likhtman and McLeish\cite{Likhtman-McLeish-2002}.
Because most of mesoscopic coarse-grained models employ segments as
fundamental units (rather than monomers), it would be convenient to use the number of
segments between entanglements $N_{e}$, instead of the molecular
weight between entanglements $M_{e}$.
(Here, we assume the Kuhn segment as the segment for coarse-grained models.)
In the same way, we employ
$N_{0}$ to express the segment number between two sequential slip-linked or
slip-springed points.
We define the average number of segments between entanglements from eq
\eqref{plateau_modulus_and_relaxation_modulus} via the classical
rubber elasticity formula (so-called the Ferry type definition)\cite{Doi-Edwards-book,Likhtman-McLeish-2002,Larson-Sridhar-Leal-McKinley-Likhtman-McLeish-2003}:
\begin{equation}
 \label{segment_number_between_entanglements_definition}
 N_{e} \equiv \frac{\rho_{0} k_{B} T}{G_{N}^{(0)}} ,
\end{equation}
where $\rho_{0}$ is the number density of segments, $k_{B}$ is the Boltzmann
constant, and $T$ is the temperature.

The segment number between entanglements $N_{e}$ defined by eq~\eqref{segment_number_between_entanglements_definition} can
be interpreted as the characteristic segment number of a model,
and naively we expect $N_{e}$ to be the input parameter.
However, as mentioned, generally $N_{0} \neq N_{e}$. Thus if we simply set $N_{0} = N_{e}$, 
the plateau modulus of a model deviates from eq~\eqref{segment_number_between_entanglements_definition}.
Although this problem itself has been recognized by researchers\cite{Masubuchi-Uneyama-2018},
as far as the authors know, the relation between $N_{0}$ and $G_{N}^{(0)}$,
or, equivalently, the relation between $N_{0}$ and $N_{e}$ is still not well understood
(except a few limited systems\cite{Steenbakkers-Tzoumanekas-Li-Liu-Kroger-Schieber-2014}).

There are various similar but different mesoscale coarse-grained
models for entangled polymers. To quantitatively
compare or connect different models,
the relation between $N_{0}$ and $G_{N}^{(0)}$ (or $N_{e}$) is important.
(For example, $N_{0}$ of one model may be quantitatively different from $N_{0}$
of another model, even if two models give the same plateau modulus.)
In this work, we theoretically calculate the relation between
$N_{0}$ and $G_{N}^{(0)}$ for various single-chain
slip-link and slip-spring type models.
We calculate the plateau modulus under the assumption that
the subchain-scale structures are fully relaxed but the slip-links
and slip-springs are not reconstructed. We show that the plateau modulus
$G_{N}^{(0)}$ depends on several parameters such as the interaction model
between slip-links and the spring constant of slip-springs.
We also perform
Monte Carlo simulations to examine the accuracy of our theoretical predictions.
Our results suggest that the various relaxation mechanisms affect the value
of $G_{N}^{(0)}$. This makes the relation between $N_{0}$ and $G_{N}^{(0)}$
(or $N_{e}$) strongly model-dependent. We examine some literature data and
discuss how we should interpret the simulation
data obtained by molecular dynamics models and mesoscopic coarse-grained models.

\section{Model}

\subsection{Equilibrium Probability Distribution}

In this work, we consider slip-link and slip-spring type models.
In these models, the entanglement effect is mimicked by introducing
transient objects which work like cross-links. These transient objects are called the slip-links or slip-springs.
Various slip-link and slip-spring type models have been proposed
for entangled polymer systems\cite{Hua-Schieber-1998,Masubuchi-Takimoto-Koyama-Ianniruberto-Greco-Marrucci-2001,Doi-Takimoto-2003,Likhtman-2005,Nair-Schieber-2006,Chappa-Morse-Zippelius-Muller-2012,Uneyama-Masubuchi-2012,Shanbhag-2019,Shanbhag-2019a}.
Although these models can reproduce characteristic dynamics of
entangled polymers, they are not equivalent. They are phenomenologically designed
based on similar but different assumptions, and thus their statistical properties
are different. As far as a model can reasonably reproduce some dynamic quantities
such as the relaxation modulus, we can employ any model. Here we do not
limit ourselves to a specific model and study some different models.

To make the analyses tractable, in this
work we limit ourselves to the single-chain type models.
In a single-chain slip-link model, we consider the dynamics of a single tagged chain and assume that
the slip-links are attached to the tagged chain. A segment to which
a slip-link is attached is spatially fixed and cannot move freely.
(This corresponds to the affine network model in the rubber elasticity theory.)
The polymer chain is allowed to slip (slide) through slip-links, and if
a slip-link reaches to the chain end, it is destroyed. To compensate
the destroyed slip-links, slip-links can be newly constructed and put on the chain.
In the rubber elasticity theory, the fluctuation of the cross-links
affects the modulus. Therefore, models with fluctuations of slip-linked points
would be preferred. To incorporate the fluctuations of slip-linked points,
we employ the slip-spring models in this work.
In the case of a single-chain slip-spring model, slip-links are replaced by slip-springs.
One end of a slip-spring is fixed in space (just like the case of a slip-link), and
another end is attached to a segment. Two ends of a slip-spring is 
connected by a harmonic spring.
(The positions of slip-springed points on the chain can fluctuate in space,
and we expect that this may mimic the non-affine network models such as the phantom
network model, in some aspects.)
The slip-springs can be dynamically destroyed
and constructed in the same way as slip-links.
These slip-link and slip-spring
models can reproduce various characteristic dynamical behaviors
of entangled polymers.
The rheological properties of a model depend both on static properties
(such as the interaction potential model) and dynamic
properties (such as the dynamic equations and transition models).
As we show in the following subsection, however,
the details of dynamics is not so important for the calculation of the
plateau modulus. We need only the equilibrium probability distribution
function to calculate the plateau modulus.

Without slip-links and slip-springs, a tagged chain is simply modeled as
an ideal Gaussian chain.
It would be fair to mention that the Gaussian chain statistics
is not always reasonable. For example, if the chain is stiff,
we should replace the Gaussian chain statistics by other statistics such
as a freely jointed chain model and models with the bending rigidity.
In this work we limit ourselves to the Gaussian chain statistics which is
the simplest and suitable to theoretical analyses.
For simplicity, in this work we assume that the total number of segments
in a chain, $N$, is given as a multiple of $N_{0}$, as $N = N_{0} Z_{0}$ with
$Z_{0}$ being a positive integer. (This assumption is not essential.
The results are not changed even if $N$ is not a multiple of $N_{0}$.
However, the expressions become a bit complicated in such a case.) $Z_{0}$ can be interpreted as the average number
of slip-linked or slip-springed subchains in the tagged chain.
Because we are
interested in the plateau modulus, we assume that $Z_{0}$ is sufficiently
large, $Z_{0} \gg 1$. (This assumption makes the statistical properties somewhat simple.)

\begin{figure}[tb!]
 \centering
 {\includegraphics[width=0.8\linewidth,clip]{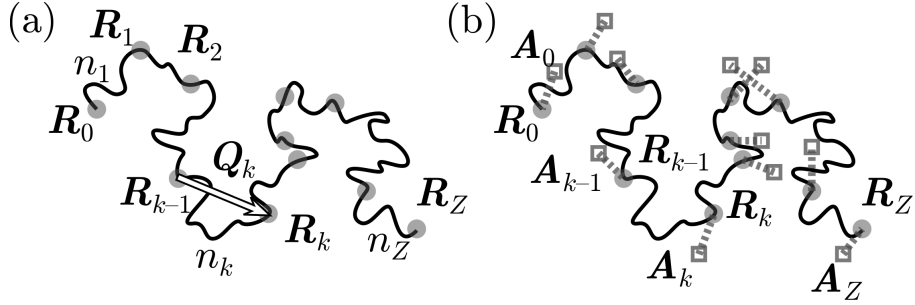}}
 \caption{Schematic images of the single-chain slip-link and slip-spring models.
 (a) The slip-link model. Slip-links (gray circles) are attached to
 a Gaussian chain (the solid black curve). The polymer chain consists of
 $Z$ subchains.
 $\bm{R}_{k}$ and $\bm{Q}_{k}$
 represent the $k$-th slip-linked point and the $k$-th bond vector, respectively.
 $n_{k}$ represents the number of segments in the $k$-th subchain.
 (b) The slip-spring model. Slip-springs (gray dotted lines) are attached to
 a Gaussian chain (the solid black curve). One end of a slip-spring is attached to
 the chain (a gray circle) whereas another end is anchored in space (a gray square).
 $\bm{A}_{k}$ represents the anchoring point of the $k$-th slip-spring.}
 \label{single_chain_model_images}
\end{figure}

We start from the single-chain slip-link models. The state of the system (the
tagged chain) is expressed by the positions of chain ends and slip-linked
points, the number of segments in slip-linked subchains, and the total
number of subchains. We express the number of subchains as $Z$ ($Z = 1,2,3,\dots$) and
the $k$-th constrained point on the chain as $\bm{R}_{k}$ ($k = 0,1,2,\dots,Z$). $k = 0$ and $Z$ correspond to the
chain ends and others correspond to slip-linked points. The number of
segments between the $(k - 1)$-th and the $k$-th points is expressed as
$n_{k}$ ($k = 1,2,\dots,Z$). An image of a tagged chain with slip-links is
depicted in Figure~\ref{single_chain_model_images}(a).
For convenience, in what follows, we use dimensionless units. We set
$\sqrt{N_{0} b^{2} / 3}$, $N_{0}$, and $k_{B} T$ as the units of length,
segment number, and energy. (This is equivalent to simply set 
$N_{0} = 1$, $b^{2} / 3 = 1$, and $k_{B} T = 1$.)

Even if we limit ourselves to the single-chain slip-link models,
there are various similar but different models.
For example, we can select the interaction potential model
for slip-links, which directly affects the statistical properties\cite{Uneyama-Masubuchi-2011}.
We start from the non-interacting slip-links.
In this work, we call this model as an ideal slip-link model.
The free energy of the ideal slip-link model is given as\cite{Schieber-2003}
\begin{equation}
 \label{free_energy_ideal_slip_link}
  \mathcal{F}_{\text{IL}}(\lbrace \bm{R}_{k} \rbrace,\lbrace n_{k}
  \rbrace,Z) =
  \sum_{k = 1}^{Z} \frac{\bm{Q}_{k}^{2}}{2 n_{k}}
  + \sum_{k = 1}^{Z} \frac{3}{2} \ln n_{k} .
\end{equation}
Here, the subscript ``IL'' represents the ideal (``I'') slip-link
(``L'') model.
For convenience, we have introduced the bond vector $\bm{Q}_{k} \equiv \bm{R}_{k} - \bm{R}_{k - 1}$
(see Figure~\ref{single_chain_model_images}(a)).
The equilibrium probability distribution for the ideal
slip-link model can be obtained by using the grand canonical type
formulation for slip-links\cite{Schieber-2003}. In the grand canonical type formulation,
slip-links are treated as a sort of ideal gas particles in one dimension.
The equilibrium probability distribution can be expressed as
\begin{equation}
 \label{probability_distribution_ideal_slip_link}
  P_{\text{eq},\text{IL}}(\lbrace \bm{R}_{k} \rbrace,\lbrace n_{k}
  \rbrace,Z) =
  \frac{\xi^{Z - 1} e^{- \xi}}{\mathcal{V}}
 \delta\bigg( Z_{0} - \sum_{k = 1}^{Z} n_{k} \bigg)
 \prod_{k = 1}^{Z} \frac{1}{(2 \pi n_{k})^{3/2}}
  \exp\bigg[ - \frac{\bm{Q}_{k}^{2}}{2 n_{k}} \bigg],
\end{equation}
where $\xi$ is the effective fugacity (activity) for slip-links, and $\mathcal{V}$ is the volume of the system.
$\xi$ is determined so that the
equilibrium average number of subchains becomes $Z_{0}$.

We also consider the single-chain slip-link models with the effective interaction between
slip-links\cite{Uneyama-Masubuchi-2011}. Although there are many possible interaction potential
models between slip-links, in this work we limit ourselves to two
special cases. One is the effective repulsion potential which cancels
the logarithmic term in eq \eqref{free_energy_ideal_slip_link}. We call
this model as the single-chain repulsive slip-link (``RL'') model. The
repulsive slip-link model can be related to the primitive chain network (PCN) model\cite{Masubuchi-Takimoto-Koyama-Ianniruberto-Greco-Marrucci-2001},
which is one of the multi-chain slip-link models.
Some properties of the repulsive slip-link model can be directly
utilized to study the PCN model.
Another is a very strong repulsive potential, with which
slip-links form a sort
of the Wigner crystal structure on the chain. We call this model as the
single-chain equidistant slip-link (``EL'') model, because the number of
segments between neighboring slip-links becomes constant 
and thus slip-links are placed equidistantly on the chain.

The free energy of the repulsive slip-link model is given as
\begin{equation}
 \label{free_energy_repulsive_slip_link}
 \mathcal{F}_{\text{RL}}(\lbrace \bm{R}_{k} \rbrace,\lbrace n_{k}
 \rbrace,Z)
  = \mathcal{F}_{\text{IL}}(\lbrace \bm{R}_{k} \rbrace,\lbrace n_{k}
 \rbrace,Z) - \sum_{k = 1}^{Z} \frac{3}{2} \ln n_{k} 
  = \sum_{k = 1}^{Z} \frac{\bm{Q}_{k}^{2}}{2 n_{k}} .
\end{equation}
Following the grand canonical type formulation, we have the formal
expression for the equilibrium probability distribution of the repulsive
slip-link model.
\begin{equation}
 \label{probability_distribution_repulsive_slip_link}
\begin{split}
  P_{\text{eq},\text{RL}}(\lbrace \bm{R}_{k} \rbrace,\lbrace n_{k}
 \rbrace,Z)
 = & \frac{\xi^{Z - 1}}{\mathcal{V} E_{5 / 2,5 / 2}(\xi)}
 \delta\bigg( Z_{0}
 - \sum_{k = 1}^{Z} n_{k} \bigg)
 \frac{1}{(2 \pi)^{3 Z / 2}}
  \exp\bigg[ - \sum_{k = 1}^{Z} \frac{\bm{Q}_{k}^{2}}{2 n_{k}}
 \bigg] \\
\end{split}
\end{equation}
where $E_{5 / 2,5 / 2}(x)$ is the generalized Mittag-Leffler function\cite{Bateman-book}
(the explicit from of $E_{5/2,5/2}(x)$ is not required for the calculations below).
As before, the effective fugacity is determined so that the average
number of subchains becomes $Z_{0}$.
Although eq \eqref{probability_distribution_repulsive_slip_link} is not
simple, if we focus on the statistics of just a single subchain in the
system, the probability distribution can be largely simplified. (See Appendix~\ref{simple_derivation_of_probability_distributions_for_single_subchain}.)
On the other hand, in the equidistant slip-link model, the equilibrium
probability distribution function becomes quite simple:
\begin{equation}
 \label{probability_distribution_equidistant_slip_link}
  P_{\text{eq},\text{EL}}(\lbrace \bm{Q}_{k} \rbrace, \lbrace n_{k}
 \rbrace, Z) =
 \frac{\delta_{Z,Z_{0}}}{\mathcal{V}}
 \bigg[\prod_{k = 1}^{Z_{0}} \delta(n_{k} - 1) \bigg]
 \frac{1}{(2 \pi)^{3 Z_{0} / 2}} \exp
 \bigg[ - \sum_{k = 1}^{Z_{0}} \frac{\bm{Q}_{k}^{2}}{2} \bigg] .
\end{equation}
Eq \eqref{probability_distribution_equidistant_slip_link} means that the
equidistantly slip-linked chain behaves as a simple bead-spring
chain. ($\delta_{Z,Z_{0}}$ in eq~\eqref{probability_distribution_equidistant_slip_link}
is the Kronecker delta which arises from the fact that $Z$ is constant
and cannot fluctuate in the equidistant slip-link model.)

In the single-chain slip-spring models, slip-linked points are allowed to fluctuate
in space around their average positions (the anchoring points).
One end of a slip-spring is attached to a slip-linked point on the chain,
and another end is anchored in space.
We express the anchoring point of the $k$-th
slip-spring as $\bm{A}_{k}$ $(k = 0,1,\dots,Z)$.
An image of a tagged chain with slip-springs is
depicted in Figure~\ref{single_chain_model_images}(b).
A slip-spring is modeled as an entropic spring which can be interpreted
as a short polymer chain. We express the number of segments for this short polymer chain as $N_{s}$.
For simplicity, we also attach
slip-springs to chain ends. (This does
not affect the plateau modulus for a sufficiently long chain with$Z_{0} \gg 1$.)

We consider three different
interaction models between slip-springs, as the case of the slip-link models explained above.
The free energy of the ideal slip-spring model is
expressed as the sum of the free energy of the ideal slip-link model and the
free energy of slip-springs:
\begin{equation}
 \label{free_energy_ideal_slip_spring}
 \mathcal{F}_{\text{IS}}(\lbrace \bm{R}_{k} \rbrace,\lbrace n_{k}
 \rbrace,Z)
 = \mathcal{F}_{\text{IL}}(\lbrace \bm{R}_{k} \rbrace,\lbrace n_{k}
 \rbrace,Z) + \sum_{k = 0}^{Z} \frac{(\bm{R}_{k} - \bm{A}_{k})^{2}}{2 \phi} .
\end{equation}
The subscript ``IS'' represents the ideal slip-spring (``S'') model.
Here, $\phi$ is the parameter which represents the effective
size of slip-springs. (In dimensional units, $\phi$ is expressed
as $\phi = N_{s} / N_{0}$.) We call $\phi$ as the 
``slip-spring size parameter'' in the followings. The free energies for
the repulsive and equidistant slip-spring (``RS'' and ``ES'')
models can be constructed in the same way.

The equilibrium probability
distributions for the single-chain slip-spring models 
can be constructed by utilizing the equilibrium probability distributions for
the single-chain slip-link models. The equilibrium probability distribution for
the anchoring points is given as a Gaussian distribution:
\begin{equation}
 P_{\text{eq},\text{S}}(\lbrace \bm{A}_{k} \rbrace | \lbrace \bm{R}_{k} \rbrace ,
  \lbrace n_{k} \rbrace, Z) = 
  \frac{1}{(2 \pi \phi)^{3 (Z + 1)/2}}
  \exp\bigg[ - \sum_{k = 0}^{Z}\frac{(\bm{R}_{k} - \bm{A}_{k})^{2}}{2 \phi} \bigg] .
\end{equation}
Here, $P_{\text{eq}}(X | Y)$ represents the conditional probability
distribution of $X$ under a given $Y$.
The subscript ``S'' represents all the three slip-springs models (``IS'',
``RS'', and ``ES'').
The equilibrium probability distribution of the
slip-spring model is thus expressed as
\begin{equation}
 \label{probability_distribution_equidistant_slip_spring}
 P_{\text{eq},\text{S}}(\lbrace \bm{R}_{k} \rbrace, \lbrace
 \bm{A}_{k} \rbrace, \lbrace n_{k} \rbrace, Z)
 =  
 \frac{1}{(2  \pi \phi)^{3 (Z + 1)/2}}
 \exp\bigg[ - \sum_{k = 0}^{Z} \frac{(\bm{R}_{k} - \bm{A}_{k})^{2}}{2 \phi} \bigg]
 P_{\text{eq},\text{L}}(\lbrace \bm{R}_{k} \rbrace, \lbrace n_{k} \rbrace, Z) .
\end{equation}
The subscript ``L'' represents the slip-link models.
(Eq \eqref{probability_distribution_equidistant_slip_spring} is common for
all the three slip-spring models examined in this work.)

\subsection{Linear Response Theory and Plateau Modulus}

We can calculate the shear relaxation modulus by evaluating the relaxation process after the
step deformation. Then the plateau modulus is determined by eq~\eqref{plateau_modulus_and_relaxation_modulus}.
Although this approach is intuitive, it requires us to
evaluate the relaxation process directly. However, the relaxation process
is generally model-dependent, and the analytic evaluation of the relaxation process is not easy.
On the other hand, according to the linear response theory,
the shear relaxation modulus of a given model can be evaluated systematically by evaluating the
correlation function\cite{Evans-Morris-book}. The standard framework of the linear response
theory is universal and we expect that we can proceed the calculation
without knowing the details of the relaxation process.
In this subsection, therefore, we derive the expression of the plateau moduli for the
single-chain slip-link and slip-spring models from the view point of the
linear response theory.

The linear response theory claims that the response function of a
physical quantity can be expressed by using the correlation function of that
physical quantity and another physical quantity which is conjugate to the applied
external field. (In the case of the relaxation modulus, both of them are the shear stress.)
For the single-chain slip-link models, the linear response theory gives
\begin{equation}
 \label{linear_response_formula_slip_link}
 G(t) = \nu_{0} \langle \hat{\sigma}_{xy}(t) \hat{\sigma}_{xy}(0) \rangle_{\text{eq}} .
\end{equation}
where $\nu_{0} \equiv \rho_{0} / Z_{0}$ is the average subchain number density and $\hat{\sigma}_{xy}(t)$ is
the $xy$-component of the single-chain stress tensor at time $t$. $\langle \dots
\rangle_{\text{eq}}$ means the statistical average in equilibrium.
(The factor $\nu_{0}$ represents the fact there are many polymer
chains in a bulk system. The single-chain model gives the
modulus just for a single chain. However, a bulk system consists of many statistically independent
chains. We should introduce the factor $\nu_{0}$ to
reproduce the relaxation modulus correctly.)
From the stress-optical rule, the single-chain stress tensor is simply given as follows:
\begin{equation}
 \label{single_chain_stress_tensor_slip_link}
  \hat{\bm{\sigma}}(t) = \sum_{k = 1}^{Z(t)}
  \frac{\bm{Q}_{k}(t) \bm{Q}_{k}(t)}{n_{k}(t)} .
\end{equation}
Here, the stress-optical coefficient is taken to be unity.
Now we consider to calculate the plateau moduli of the slip-link models by eq
\eqref{linear_response_formula_slip_link}.
Clearly it is not easy to calculate $G(t)$ exactly. (It requires the
detailed information of the dynamics.) Nevertheless, we are able to obtain
accurate approximate expressions under several conditions.
From eq~\eqref{plateau_modulus_and_relaxation_modulus}, it is sufficient
for us to calculate the value of $G(t)$ only around $t = \tau_{e}$.
Besides, at this time scale, $G(t)$ is expected to be almost constant.
At the time scale of $t \approx \tau_{e}$, we can reasonably assume that
the slip-links are not constructed nor destroyed.
Thus here we assume that the positions of slip-links are not changed.
However, the transport of segments between neighboring subchains occurs
even at the relatively short time scale.
Thus we expect that the segment numbers in subchains are locally equilibrated at this time scale.
We consider that only the segment numbers are equilibrated and
other variables are unchanged at $t \approx \tau_{e}$.
Then we can set $\bm{R}_{k}(t) =
\bm{R}_{k}(0) = \bm{R}_{k}$ and $Z(t) = Z(0) = Z$ in eq~\eqref{linear_response_formula_slip_link},
and assume that $\lbrace n_{k}(0) \rbrace$ and $\lbrace n_{k}(\tau_{e}) \rbrace$ are
sampled independently from the local equilibrium distribution.
Under these assumptions, the plateau modulus can be approximately expressed as
\begin{equation}
 \label{linear_response_formula_slip_link_approx_local_equilibrium}
  \begin{split}
   G_{N,\text{L}}^{(0)}
   & \approx \nu_{0}
   \sum_{Z = 0}^{\infty} \int d\lbrace \bm{R}_{k} \rbrace \, \bigg[
   \int d\lbrace n_{k} \rbrace
   \sum_{k = 1}^{Z} \frac{Q_{k,x}
   Q_{k,y}}{n_{k}} P_{\text{eq},\text{L}}(\lbrace n_{k} \rbrace | \lbrace
   \bm{R}_{k} \rbrace, Z) \bigg]^{2}
   P_{\text{eq},\text{L}}(\lbrace \bm{R}_{k} \rbrace, Z) .
  \end{split}
\end{equation}

For the single-chain slip-spring models, we should take into account
the contribution of slip-springs (the virtual stress).
The linear response formula
becomes as follows\cite{Ramirez-Sukumaran-Likhtman-2007,Uneyama-2011}:
\begin{equation}
 \label{linear_response_formula_slip_spring}
 G(t) = \nu_{0} \langle \hat{\sigma}_{xy}(t) \hat{\sigma}_{xy}(0)
 \rangle_{\text{eq}}
 + \nu_{0} \langle \hat{\sigma}_{xy}(t) \hat{\sigma}_{xy}^{(v)}(0)
 \rangle_{\text{eq}} .
\end{equation}
Here $\hat{\sigma}_{xy}^{(v)}$ is the virtual stress tensor by slip-springs:
\begin{equation}
 \label{single_chain_virtual_stress_tensor_slip_spring}
  \hat{\bm{\sigma}}^{(v)}(t) = \sum_{k = 0}^{Z(t)} 
  \frac{(\bm{R}_{k}(t) - \bm{A}_{k}(t)) (\bm{R}_{k}(t) - \bm{A}_{k}(t))}{\phi} .
\end{equation}
It would be fair to mention that the definition of the
stress in the slip-spring models is not fully clear. From the
linear response theory\cite{Uneyama-2011}, $\hat{\sigma}_{xy} + \hat{\sigma}_{xy}^{(v)}$ is
conjugate to the applied strain, and should be employed as the stress.
However, $\hat{\sigma}_{xy} + \hat{\sigma}^{(v)}_{xy}$ is generally not consistent with the stress-optical rule.
Thus, from the stress-optical rule, we should employ $\hat{\sigma}_{xy}$ as the stress tensor
(as the case of the slip-link models). Here we assume that the stress-optical rule holds
and employ $\hat{\sigma}_{xy}$ as the stress of the slip-spring models. This gives eq~\eqref{linear_response_formula_slip_spring}.
At $t \approx \tau_{e}$, we expect that the positions $\lbrace
\bm{R}_{k} \rbrace$ and segment numbers $\lbrace n_{k} \rbrace$ are
locally equilibrated while other variables are unchanged.
Then we have the following approximate expression for the plateau
modulus:
\begin{equation}
 \label{linear_response_formula_slip_spring_approx_local_equilibrium}
  \begin{split}
   G_{N,\text{S}}^{(0)}
   & \approx \nu_{0}
   \sum_{Z = 0}^{\infty} \int d\lbrace \bm{A}_{k} \rbrace \,
     \bigg[
   \int d\lbrace \bm{R}_{k} \rbrace
   d\lbrace n_{k} \rbrace \,
   \sum_{k = 1}^{Z} \frac{Q_{k,x}
   Q_{k,y}}{n_{k}}
   P_{\text{eq},\text{S}}(\lbrace \bm{R}_{k} \rbrace, \lbrace
   n_{k} \rbrace | \lbrace \bm{A}_{k} \rbrace, Z) \bigg]
   \\
   & \qquad \times
   \bigg[
   \int d\lbrace \bm{R}_{k} \rbrace
   d\lbrace n_{k} \rbrace \,
   \bigg[ \sum_{k = 1}^{Z} \frac{Q_{k,x}
   Q_{k,y}}{n_{k}}
   + \sum_{k = 0}^{Z} \frac{(R_{k,x} - A_{k,x})
   (R_{k,y} - A_{k,y})}{\phi} \bigg] \\
   & \qquad \times    P_{\text{eq},\text{S}}(\lbrace \bm{R}_{k} \rbrace, \lbrace n_{k} \rbrace | \lbrace \bm{A}_{k} \rbrace, Z)
   \bigg] P_{\text{eq},\text{S}}(\lbrace \bm{A}_{k} \rbrace, Z) .
  \end{split}
\end{equation}
From the expressions shown above, we can calculate
the plateau moduli of slip-link and slip-spring models only from
the local equilibrium probability distributions.

\section{Theory}

\subsection{Single-Subchain Approximation}

To proceed the calculation and obtain the explicit expressions for the plateau
moduli of the slip-link and slip-spring models, we employ the single-subchain approximation. 
We consider only a single tagged subchain in a chain. Such an approximation
enables us to analytically calculate various statistical quantities including the plateau
modulus.

\begin{figure}[tb!]
 \centering
 {\includegraphics[width=0.6\linewidth,clip]{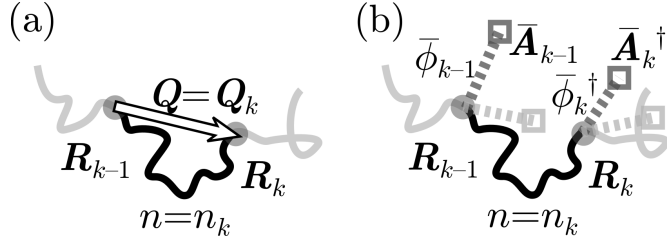}}
 \caption{The single-subchain approximation for the single-chain slip-link and slip-spring models.
 (a) The slip-link model. The statistics of a single subchain (the black solid curve)
 can be expressed as a function of the bond vector $\bm{Q}$ and the segment number $n$.
 Other subchains (solid light gray curves) are not explicitly considered.
 (b) The slip-spring model.
 The anchoring points and slip-spring size parameters of two slip-springs
 attached to the tagged subchain are expressed by using effective anchoring points
 ($\bar{\bm{A}}_{k - 1}$ and $\bar{\bm{A}}_{k}^{\dagger}$, gray squares)
 and effective slip-spring size parameters ($\bar{\phi}_{k - 1}$ and $\bar{\phi}_{k}^{\dagger}$).
 The effective anchoring points and slip-spring size parameters are not
 the same as original anchoring points (light gray circles) and the original slip-spring size parameter.
 }
 \label{single_subchain_approximation}
\end{figure}

For the single-chain slip-link models, the state of a single subchain can be
fully expressed by the number of segments in the subchain and the 
bond vector of the subchain.
We consider the $k$-th subchain as a tagged subchain and express $n =
n_{k}$ and $\bm{Q} = \bm{Q}_{k}$ (as schematically shown in Figure~\ref{single_subchain_approximation}(a)).
The probability distribution function for a single subchain,
$P_{\text{eq}}(\bm{Q},n)$, is already obtained for several slip-link models.\cite{Uneyama-Masubuchi-2011}
The segment number distribution function $P_{\text{eq}}(n)$
and the bond vector distribution function $P_{\text{eq}}(\bm{Q})$ are
calculated straightforwardly from $P_{\text{eq}}(\bm{Q},n)$.
We show the simple derivation which gives the same result as the previous work\cite{Uneyama-Masubuchi-2011}
in Appendix~\ref{simple_derivation_of_probability_distributions_for_single_subchain}.
In the slip-link models,
the segment number distribution function is given as follows:
\begin{equation}
 \label{segment_number_distribution_single_subchain}
 P_{\text{eq},\text{L}}(n) =
  \begin{cases}
   e^{-n} & (\text{ideal}) , \\
   \displaystyle \frac{25}{6} \sqrt{\frac{10}{\pi}} n^{3/2}
   e^{- 5 n / 2} & (\text{repulsive}) , \\
   \delta(n - 1) & (\text{equidistant}) .
  \end{cases} 
\end{equation}
The bond vector distribution under a
given $n$ is simply given as the Gaussian distribution:
\begin{equation}
 P_{\text{eq},\text{L}}(\bm{Q} | n) = \frac{1}{(2 \pi
  n)^{3/2}}  e^{- \bm{Q}^{2} / 2 n } .
\end{equation}
The joint probability distribution function for the segment number and the bond
vector is $P_{\text{eq},\text{L}}(\bm{Q},n) = P_{\text{eq},\text{L}}(\bm{Q}|n) P_{\text{eq},\text{L}}(n)$.
The bond vector distribution function is calculated as $P_{\text{eq},\text{L}}(\bm{Q}) = \int dn \, P_{\text{eq},\text{L}}(\bm{Q},n)$.
The result is:
\begin{equation}
 \label{bond_vector_distribution_single_subchain}
 P_{\text{eq},\text{L}}(\bm{Q}) =
  \begin{cases}
   \displaystyle \frac{1}{2 \pi |\bm{Q}|}
 e^{- \sqrt{2} |\bm{Q}|} & (\text{ideal}), \\
   \displaystyle \frac{25}{6 \pi^{2}} |\bm{Q}| K_{1}(\sqrt{5} |\bm{Q}|) & (\text{repulsive}), \\
   \displaystyle \frac{1}{(2 \pi)^{3/2}} e^{- \bm{Q}^{2} / 2} & (\text{equidistant}).
  \end{cases}
\end{equation}
Here, $K_{1}(x)$ is the first order modified Bessel function of the second kind\cite{Abramowitz-Stegun-book,NIST-handbook}.
From the Bayes' theorem, the conditional probability distribution function
of $n$ under given $\bm{Q}$ is
calculated from eqs \eqref{segment_number_distribution_single_subchain}
and \eqref{bond_vector_distribution_single_subchain}, as
$P_{\text{eq},\text{L}}(n|\bm{Q}) = P_{\text{eq},\text{L}}(\bm{Q},n) / P_{\text{eq},\text{L}}(\bm{Q})$.
Thus we have
\begin{equation}
 \label{conditional_segment_number_distribution_single_subchain}
 P_{\text{eq},\text{L}}(n | \bm{Q}) =
  \begin{cases}
   \displaystyle \frac{1}{\sqrt{2 \pi}}
  \frac{| \bm{Q} |}{ n^{3/2}} 
 \exp\bigg(- \frac{\bm{Q}^{2}}{2 n} - n
 +  \sqrt{2} | \bm{Q} | \bigg) & (\text{ideal}) , \\
   \displaystyle \frac{\sqrt{5}}{2}
  \frac{1}{| \bm{Q} |
  K_{1}(\sqrt{5} | \bm{Q} |)} 
  \exp \bigg( - \frac{\bm{Q}^{2}}{2 n}
  - \frac{5 n}{2} \bigg) & (\text{repulsive}) , \\
   \delta(n - 1) & (\text{equidistant}) .
  \end{cases}
\end{equation}

Finally, by using the equilibrium
probability distribution functions
\eqref{bond_vector_distribution_single_subchain} and
\eqref{conditional_segment_number_distribution_single_subchain}, the plateau modulus
of the single-chain slip-link model becomes
\begin{equation}
 \label{linear_response_formula_slip_link_approx_single_subchain}
  \begin{split}
   G_{N,\text{L}}^{(0)}
   & \approx \nu_{0}
   \sum_{Z = 1}^{\infty}  Z \bigg[ \int d\bm{Q} \, \bigg[
   \int dn \,
   \frac{1}{n}
   P_{\text{eq},\text{L}}(n | \bm{Q}) \bigg]^{2}
   Q^{2}_{x} Q^{2}_{y} P_{\text{eq},\text{L}}(\bm{Q})
   \bigg] P_{\text{eq},\text{L}}(Z) \\
   & = G_{0} \int d\bm{Q} \, \bigg[
   \int dn \,
   \frac{1}{n}
   P_{\text{eq},\text{L}}(n | \bm{Q}) \bigg]^{2}
   Q_{x}^{2} Q_{y}^{2} P_{\text{eq},\text{L}}(\bm{Q}) .
  \end{split}
\end{equation}
Here we have defined the characteristic modulus $G_{0}$ as $G_{0} \equiv \nu_{0} Z_{0} = \rho_{0}$
(This characteristic modulus is common for all the single-chain slip-link and slip-spring models,
as shown in Appendix~\ref{characteristic_modulus}.)
In dimensional units, $G_{0}$ is expressed as $G_{0} = \nu_{0} k_{B} T N /
N_{0} = \rho_{0} k_{B} T / N_{0}$. Therefore, $G_{0}$ corresponds to the modulus of an ideal rubber
with the average subchain segment number $N_{0}$.
In eq~\eqref{linear_response_formula_slip_link_approx_single_subchain},
the integral over $n$ explicitly depends on the distribution function $P_{\text{eq},\text{L}}(n|\bm{Q})$.
Therefore, models with different distribution functions
have different plateau moduli.

Next we consider the single-chain slip-spring models. In the case of the slip-spring
models, we need to specify the anchoring points in addition to 
the number of segments in the subchain and the bond vector.
Under the single-subchain approximation, we cannot simply utilize
the anchoring points and the slip-spring size parameters for the full chain model.
The spatial fluctuation
of a slip-linked point is affected by two subchains
and one slip-spring connected to the point. If we simply cut the subchains connected to
a tagged subchain, the fluctuation is affected only by one subchain and one slip-spring.
This situation is clearly different from that of the original tagged subchain.
To recover the correct fluctuation, we should employ the effective anchoring points and
the effective slip-spring size parameters.
 We consider the $k$-th subchain as a tagged subchain, as before, and
express the effective anchoring points for the $(k - 1)$-th and the $k$-th
anchoring points as $\bar{\bm{A}}_{k - 1}$ and $\bar{\bm{A}}_{k}^{\dagger}$,
respectively. Also, we express the effective slip-spring size parameters
for the $(k - 1)$-th and $k$-th slip-springs as $\bar{\phi}_{k - 1}$ and $\bar{\phi}_{k}^{\dagger}$.
The schematic image of the single-subchain approximation for the slip-spring model
is shown in Figure~\ref{single_subchain_approximation}(b).
The effective anchoring points and the effective slip-spring size parameters
are obtained by integrating-out the degrees of freedom of other subchains.
Then we can express the equilibrium probability distribution of a single subchain
as $P_{\text{eq},\text{S}}(\bm{R}_{k - 1},\bm{R}_{k},n_{k},\lbrace \bm{A}_{k} \rbrace)$.
Unlike the case of the slip-link model, the bond vector can be
equilibrated at the entanglement time scale in the slip-spring model.
To calculate the plateau modulus, we consider that the number of segments and the
bond vector obey the equilibrium distribution under a given configuration of the anchoring point,
$P_{\text{eq},\text{S}}(\bm{R}_{k - 1}, \bm{R}_{k} ,n_{k} | \lbrace \bm{A}_{k} \rbrace)$.
We have the following expression for the plateau modulus:
\begin{equation}
 \label{linear_response_formula_slip_spring_approx_single_subchain}
  \begin{split}
   G_{N,\text{S}}^{(0)}
   & \approx G_{0} \int d\lbrace \bm{A}_{k} \rbrace \,   P_{\text{eq},\text{S}}(\lbrace \bm{A}_{k} \rbrace)
   \bigg[  \int dn d\bm{R}_{k - 1}d\bm{R}_{k} \,
   \frac{Q_{k,x} Q_{k,y}}{n_{k}}
   P_{\text{eq},\text{S}}(\bm{R}_{k - 1}, \bm{R}_{k}, n_{k} | \lbrace \bm{A}_{k} \rbrace) \bigg] \\
   & \qquad \times \bigg[ \int dn d\bm{R}_{k - 1}d\bm{R}_{k} \,
   \bigg[ \frac{Q_{k,x} Q_{k,y}}{n_{k}} 
   + \frac{2 (R_{k - 1,x} - \bar{A}_{k - 1,x}) (R_{k-1,y}- \bar{A}_{k - 1,y})}{\bar{\phi}_{k - 1}} \\
   & \qquad 
   + \frac{2 (R_{k,x} - \bar{A}_{k,x}^{\dagger}) (R_{k,y} - \bar{A}_{k,y}^{\dagger})}{\bar{\phi}_{k}^{\dagger}}
   \bigg]
   P_{\text{eq},\text{S}}(\bm{R}_{k - 1}, \bm{R}_{k}, n_{k} | \lbrace \bm{A}_{k} \rbrace) \bigg] .
  \end{split}
\end{equation}

To calculate the plateau modulus by eq~\eqref{linear_response_formula_slip_spring_approx_single_subchain},
we should evaluate the integrals over $n_{k}$, $\bm{R}_{k - 1}$, and
$\bm{R}_{k}$. Unfortunately,
even under a single subchain approximation, such integrals cannot be
easily evaluated in general. For the equidistant slip-spring model,
however, the expression reduces to be simple. Since the number of segments is
constant ($n_{k} = 1$), eq~\eqref{linear_response_formula_slip_spring_approx_single_subchain}
reduces to
\begin{equation}
 \label{linear_response_formula_equidistant_slip_spring_approx_single_subchain}
  \begin{split}
   G_{N,\text{ES}}^{(0)}
   & \approx G_{0} \int d\lbrace \bm{A}_{k} \rbrace \,   P_{\text{eq},\text{ES}}(\lbrace \bm{A}_{k} \rbrace)
   \bigg[ \int d\bm{R}_{k - 1}d\bm{R}_{k} \,
  Q_{k,x} Q_{k,y}
   P_{\text{eq},\text{ES}}(\bm{R}_{k - 1}, \bm{R}_{k} | \lbrace \bm{A}_{k} \rbrace) \bigg] \\
   & \qquad \times \bigg[ \int d\bm{R}_{k - 1}d\bm{R}_{k} \,
  \bigg[ Q_{k,x} Q_{k,y}
   + \frac{(R_{k - 1,x} - \bar{A}_{k - 1,x}) (R_{k-1,y}- \bar{A}_{k - 1,y})}{\bar{\phi}_{k - 1}} \\
   & \qquad 
   + \frac{(R_{k,x} - \bar{A}_{k,x}^{\dagger}) (R_{k,y} - \bar{A}_{k,y}^{\dagger})}{\bar{\phi}_{k}^{\dagger}}
   \bigg]
   P_{\text{eq},\text{ES}}(\bm{R}_{k - 1}, \bm{R}_{k} | \lbrace \bm{A}_{k} \rbrace) \bigg] .
  \end{split}
\end{equation}
Here, we note that the equidistant slip-spring model looks similar to the rubber elasticity model by
Rubinstein and Panyukov\cite{Rubinstein-Panyukov-1997,Rubinstein-Panyukov-2002}.
Eq~\eqref{linear_response_formula_equidistant_slip_spring_approx_single_subchain} corresponds to
the comb polymer model in the Rubinstein-Panyukov model.
For the ideal and repulsive slip-spring models, we employ the decoupling
approximation for $n_{k}$, and $\bm{R}_{k - 1}$ and $\bm{R}_{k}$. Under the decoupling approximation,
we calculate the contributions of the integral over $n$ and that over $\bm{R}_{k - 1}$ and $\bm{R}_{k}$,
separately. The former is the same as the case of the slip-link model,
and the latter is the same as the equidistant slip-spring model. Therefore,
we have the following approximate expression:
\begin{equation}
 \label{linear_response_formula_slip_spring_approx_single_subchain_decoupling}
  \frac{G_{N,\text{S}}^{(0)}}{G_{0}}
  \approx \frac{G_{N,\text{ES}}^{(0)}}{G_{0}}
  \frac{G_{N,\text{L}}^{(0)}}{G_{0}} .
\end{equation}

To obtain the explicit expression of the plateau modulus for the equidistant
slip-spring model, we need the equilibrium distribution function 
for the single subchain with the effective anchoring points and the
effective slip-spring sizes.
Although the exact expressions become quite complicated, we can reasonably
approximate them by the following forms:
\begin{align}
 \label{effective_slip_spring_size_single_subchain}
 \bar{\phi}_{k - 1} & \approx \bar{\phi}_{k}^{\dagger}
  \approx \bar{\phi}_{\infty} \equiv \frac{- 1 + \sqrt{1 + 4 \phi}}{2}, \\
 \label{effective_anchoring_point_left_single_subchain}
 \bar{\bm{A}}_{k - 1} & \approx
 \frac{\bar{\phi}_{\infty}}{\phi} \sum_{j = 0}^{\infty} 
 \left(1 - \frac{\bar{\phi}_{\infty}}{\phi}\right)^{j} \bm{A}_{k - 1 - j}, \\
 \label{effective_anchoring_point_right_single_subchain}
 \bar{\bm{A}}_{k}^{\dagger} & \approx
 \frac{\bar{\phi}_{\infty}}{\phi} \sum_{j = 0}^{\infty}
  \left(1 - \frac{\bar{\phi}_{\infty}}{\phi}\right)^{j} \bm{A}_{k + j} .
\end{align}
The detailed derivations of eqs~\eqref{effective_slip_spring_size_single_subchain}-\eqref{effective_anchoring_point_right_single_subchain}
are shown in Appendix~\ref{single_subchain_statistics_in_equidistant_slip_spring_model}.
With the effective anchoring points and the effective slip-spring size
parameters by eqs~\eqref{effective_slip_spring_size_single_subchain}-\eqref{effective_anchoring_point_right_single_subchain},
the equilibrium distribution function for the single subchain becomes
\begin{equation}
 \label{probability_distribution_slip_spring_single_subchain}
  \begin{split}
  P_{\text{eq},\text{ES}}(\bm{R}_{k - 1},\bm{R}_{k} | \lbrace \bm{A}_{k} \rbrace)
 & = \frac{(1 + 2 \bar{\phi}_{\infty})^{3/2}}{(2 \pi \bar{\phi}_{\infty})^{3}} \exp
  \bigg[ - \frac{(\bm{R}_{k} - \bm{R}_{k - 1})^{2}}{2} 
   - \frac{(\bm{R}_{k - 1} - \bar{\bm{A}}_{k - 1})^{2}}{2 \bar{\phi}_{\infty}} \\
 & \qquad - \frac{(\bm{R}_{k} - \bar{\bm{A}}^{\dagger}_{k})^{2}}{2 \bar{\phi}_{\infty}}
   + \frac{(\bar{\bm{A}}_{k} - \bar{\bm{A}}^{\dagger}_{k})^{2}}{1 + 2 \bar{\phi}_{\infty}}
\bigg] .
  \end{split}
\end{equation}

\subsection{Slip-Link Models}
\label{slin_link_type_models}

We calculate the plateau moduli of the single-chain slip-link models.
The simplest case is the equidistant slip-link model.
In the case of the equidistant slip-link model, from eq~\eqref{conditional_segment_number_distribution_single_subchain},
the integral in eq~\eqref{linear_response_formula_slip_link_approx_single_subchain} is trivial:
\begin{equation}
  \int dn \, \frac{1}{n} P_{\text{eq},\text{EL}}(n | \bm{Q})
   = 1 .
\end{equation}
Then we have the following explicit form for the plateau modulus:
\begin{equation}
 \frac{G_{N,\text{EL}}^{(0)}}{G_{0}}
  \approx \int d\bm{Q} \,
  Q_{x}^{2} Q_{y}^{2} P_{\text{eq},\text{EL}}(\bm{Q})
  = 1 .
\end{equation}
The equidistant slip-link model has no fluctuation and thus this result
is physically reasonable. Roughly speaking, this result corresponds
to the pure reptation model (without the relaxation by the longitudinal motion).

For the ideal or repulsive slip-link models, explicit expressions for
the integral in eq~\eqref{linear_response_formula_slip_link_approx_single_subchain} become
a bit complicated. For the ideal slip-link model, we have
\begin{equation}
 \begin{split}
  \int dn \, \frac{1}{n} \, P_{\text{eq},\text{IL}}(n | \bm{Q})
  & =
  \int_{0}^{\infty} dn \, \frac{1}{n}
  \frac{1}{\sqrt{2 \pi}}
  \frac{|\bm{Q}|}{n^{3/2}}
  \exp \bigg( - \frac{\bm{Q}^{2}}{2 n}
  - n + \sqrt{2} | \bm{Q} | \bigg)
  \\
  & = \frac{1}{\bm{Q}^{2}} +
  \frac{\sqrt{2}}{ | \bm{Q} |} ,
 \end{split}
\end{equation}
and thus the plateau modulus is approximately expressed as
\begin{equation}
 \begin{split}
  \frac{G_{N,\text{IL}}^{(0)}}{G_{0}}
  & \approx \int d\bm{Q} \,
 \bigg( \frac{1}{\bm{Q}^{2}} +
  \frac{\sqrt{2}}{ | \bm{Q} | }
 \bigg)^{2} Q_{x}^{2}
  Q_{y}^{2} P_{\text{eq},\text{IL}}(\bm{Q}) .
 \end{split}
\end{equation}
The integral over $\bm{Q}$ can be simplified by using the polar coordinates,
($Q_{x} = Q \cos \theta \cos \varphi$, $Q_{y} = Q \cos \theta \sin
\varphi$, and $Q_{z} = Q \sin \theta$):
\begin{equation}
 \begin{split}
  \frac{G_{N,\text{IL}}^{(0)}}{G_{0}}
  & \approx \int_{0}^{\infty} dQ \int_{0}^{2 \pi} d\theta
  \int_{-\pi / 2}^{\pi / 2} d\varphi \,
  Q^{2} \cos\varphi
  \Big[ (1 +
  \sqrt{2} Q
 )^{2} \cos^{4}\varphi \sin^{2}\theta \cos^{2}\theta \Big] 
  \frac{1}{2 \pi Q} e^{- \sqrt{2} Q} \\
  & = \frac{2}{15} \int_{0}^{\infty} dQ
 \,
 [ Q +
  2 \sqrt{2} Q^{2}
  + 2 Q^{3}
 ] e^{- \sqrt{2} Q} = \frac{11}{15}.
 \end{split}
\end{equation}
Then we have the following simple expression for the plateau modulus
of the ideal slip-link model:
\begin{equation}
 \frac{G_{N,\text{IL}}^{(0)}}{G_{0}} = \frac{11}{15} \approx 0.7333 .
\end{equation}
Unlike the case of the equidistant slip-link model, the
plateau modulus of the ideal slip-link model is lower than the
characteristic modulus $G_{0}$. This can be interpreted as the effect of
the fluctuation of the segment number.

We consider the repulsive slip-link model. This model has
probability distributions less simpler than other slip-link models.
The integral in eq~\eqref{linear_response_formula_slip_link_approx_single_subchain} becomes
\begin{equation}
 \begin{split}
  \int dn \, \frac{1}{n} P_{\text{eq},\text{RL}}(n | \bm{Q})
  & =   \frac{\sqrt{5}}{2}
  \frac{1}{|\bm{Q}|
  K_{1}(\sqrt{5} | \bm{Q} |)} 
\int_{0}^{\infty} dn \, \frac{1}{n}
  \exp \bigg( - \frac{\bm{Q}^{2}}{2 n}
  - \frac{5 n}{2} \bigg) \\
  & = - 
  \frac{2}{ | \bm{Q} |
  K_{1}(\sqrt{5} | \bm{Q} |)} \frac{\partial}{\partial (\bm{Q}^{2})}
  \Big[ | \bm{Q} |
  K_{1}(\sqrt{5} | \bm{Q} |)  \Big]
  \\
  & = \frac{\sqrt{5}}{|\bm{Q}|} \frac{K_{0}(\sqrt{5}
  |\bm{Q}|)}{K_{1}(\sqrt{5} |\bm{Q}|)}.
 \end{split}
\end{equation}
As the case of the ideal slip-link model, the polar coordinates are convenient to calculate the plateau modulus.
The plateau modulus becomes
\begin{equation}
 \begin{split}
  \frac{G_{N,\text{RL}}^{(0)}}{G_{0}}
  & \approx \int d\bm{Q} \,
  \bigg[ \frac{\sqrt{5}}{|\bm{Q}|}
  \frac{K_{0}(\sqrt{5} |\bm{Q}|)}{K_{1}(\sqrt{5} |\bm{Q}|)} 
  \bigg]^{2}
  Q_{x}^{2} Q_{y}^{2} P_{\text{eq},\text{RL}}(\bm{Q}) \\
  & = \frac{1}{6 \pi^{2}} \int_{0}^{\infty} d\zeta \int_{0}^{\infty} d\theta 
  \int_{-\pi / 2}^{\pi / 2} d\varphi  \, 
  \frac{K_{0}^{2}({\zeta})}{K_{1}(\zeta)}
  \zeta^{5} \cos^{5} \varphi \cos^{2} \theta \sin^{2} \theta
 \\
  & = \frac{2}{45 \pi}
  \int_{0}^{\infty} d\zeta \,
  \zeta^{5}  \frac{K_{0}^{2}({\zeta})}{K_{1}(\zeta)},
 \end{split}
\end{equation}
where we have set $\zeta = \sqrt{5} |\bm{Q}|$.
By numerically evaluating the integral over $\zeta$, we have the following expression
for the plateau modulus for the repulsive slip-link model:
\begin{equation}
 \frac{G_{N,\text{RL}}^{(0)}}{G_{0}} \approx 0.8214 .
\end{equation}
Thus we find that the plateau modulus of the repulsive slip-link model is smaller
than that of the equidistant model yet is larger than that of the ideal slip-link
model. The fluctuation of the segment number in the repulsive slip-link
model is relatively suppressed (by the repulsive interaction), and thus the
resulting plateau modulus becomes larger than that of the ideal
slip-link model.

\subsection{Slip-Spring Models}
\label{slip_spring_type_models}

As we mentioned, we calculate the plateau modulus of the slip-spring
models by utilizing the decoupling approximation (eq~\eqref{linear_response_formula_slip_spring_approx_single_subchain_decoupling}).
Therefore, what we need
to calculate here is the plateau modulus of the equidistant slip-spring model by eq~\eqref{linear_response_formula_equidistant_slip_spring_approx_single_subchain}.
The integrals over $\bm{R}_{k-1} $ and $\bm{R}_{k}$ in eq~\eqref{linear_response_formula_equidistant_slip_spring_approx_single_subchain} become as follows:
\begin{equation}
 \begin{split}
  & \int d\bm{R}_{k - 1}d\bm{R}_{k} \,
  Q_{k,x} Q_{k,y}
   P_{\text{eq},\text{ES}}(\bm{R}_{k - 1}, \bm{R}_{k} | \lbrace \bm{A}_{k} \rbrace) \\
  & = \frac{ ( \bar{A}_{k,x}^{\dagger} - \bar{A}_{k - 1,x})  ( \bar{A}_{k,y}^{\dagger} - \bar{A}_{k - 1,y})}{(1 + 2 \bar{\phi}_{\infty})^{2}} ,
 \end{split}
\end{equation}
\begin{equation}
 \begin{split}
  & \int d\bm{R}_{k - 1}d\bm{R}_{k} \,
  \frac{(R_{k - 1,x} - \bar{A}_{k - 1,x}) (R_{k-1,y}- \bar{A}_{k - 1,y})}{\bar{\phi}_{\infty}}
   P_{\text{eq},\text{ES}}(\bm{R}_{k - 1}, \bm{R}_{k} | \lbrace \bm{A}_{k} \rbrace) \\
  & = \int d\bm{R}_{k - 1}d\bm{R}_{k} \,
  \frac{(R_{k,x} - \bar{A}_{k,x}^{\dagger}) (R_{k,y}- \bar{A}_{k,y}^{\dagger})}{\bar{\phi}_{\infty}}
   P_{\text{eq},\text{ES}}(\bm{R}_{k - 1}, \bm{R}_{k} | \lbrace \bm{A}_{k} \rbrace) \\
  & = \frac{ \bar{\phi}_{\infty} ( \bar{A}_{k,x}^{\dagger} - \bar{A}_{k - 1,x})  ( \bar{A}_{k,y}^{\dagger} - \bar{A}_{k - 1,y})}{(1 + 2 \bar{\phi}_{\infty})^{2}} .
 \end{split}
\end{equation}
Then eq~\eqref{linear_response_formula_equidistant_slip_spring_approx_single_subchain}
can be rewritten in a simple form as
\begin{equation}
 \label{linear_response_formula_equidistant_slip_spring_approx_single_subchain_modified}
  \begin{split}
   \frac{G_{N,\text{ES}}^{(0)}}{G_{0}}
   & \approx \frac{1}{(1 + 2 \bar{\phi}_{\infty})^{3}} \sum_{j, k = 1}^{Z_{0}}
   \int d\lbrace \bm{A}_{k} \rbrace \, 
   ( \bar{A}_{k,x}^{\dagger} - \bar{A}_{k - 1,x})^{2} ( \bar{A}_{k,y}^{\dagger} - \bar{A}_{k - 1,y})^{2} P_{\text{eq},\text{ES}}(\lbrace \bm{A}_{k} \rbrace)\\
   & = \frac{1}{(1 + 2 \bar{\phi}_{\infty})^{3}}
 \langle ( \bar{A}_{k,x}^{\dagger} - \bar{A}_{k - 1,x})^{2}\rangle_{\text{eq}}^{2}.
  \end{split}
\end{equation}
Thus we find that we can calculate
the plateau modulus of the equidistant slip-spring model once
the explicit expression for the variance for $\bar{A}_{k,x}^{\dagger} - \bar{A}_{k - 1,x}$ is obtained.

The vector $\bar{\bm{A}}_{k}^{\dagger} - \bar{\bm{A}}_{k}$ can be rewritten
in terms of the bond vector for the anchoring point, $\bm{U}_{k} \equiv \bm{A}_{k} - \bm{A}_{k - 1}$, as
\begin{equation}
 \label{effective_anchoring_point_bond_vector_equidistant_slip_spring}
 \begin{split}
  \bar{\bm{A}}_{k}^{\dagger} - \bar{\bm{A}}_{k - 1}
  & = \frac{\bar{\phi}_{\infty}}{\phi} \sum_{j = 0}^{\infty}
  \left(1 - \frac{\bar{\phi}_{\infty}}{\phi} \right)^{j}
  (\bm{A}_{k + j} - \bm{A}_{k - 1 - j}) \\
  & = \frac{\bar{\phi}_{\infty}}{\phi} \sum_{j = 0}^{\infty}
  \left(1 - \frac{\bar{\phi}_{\infty}}{\phi} \right)^{j}
  (\bm{U}_{k + j} + \dots + \bm{U}_{k - j}) \\
  & = \sum_{j = -\infty}^{\infty}
  \left(1 - \frac{\bar{\phi}_{\infty}}{\phi} \right)^{|j|}
  \bm{U}_{k + j} .
 \end{split}
\end{equation}
The bond vectors $\lbrace \bm{U}_{k} \rbrace$ obey the Gaussian distribution,
and the mean and variance are given as
\begin{equation}
 \label{anchoring_point_bond_vector_mean_and_variance_equidistant_slip_spring}
 \langle \bm{U}_{k} \rangle_{\text{eq}} = 0, \qquad
  \langle \bm{U}_{k} \bm{U}_{j} \rangle_{\text{eq}} =
  \begin{cases}
   (1 + 2 \phi) \bm{1} & (k = j), \\
   - \phi \bm{1} & (k = j \pm 1), \\
   0 & (\text{otherwise}).
  \end{cases}
\end{equation}
Here, $\bm{1}$ represents the unit tensor.
The detailed calculations are shown in Appendix~\ref{single_subchain_statistics_in_equidistant_slip_spring_model}.
From eqs~\eqref{effective_anchoring_point_bond_vector_equidistant_slip_spring}
and \eqref{anchoring_point_bond_vector_mean_and_variance_equidistant_slip_spring}, the variance of  $\bar{A}_{k,x}^{\dagger} - \bar{A}_{k - 1,x}$ is calculated to be
\begin{equation}
 \begin{split}
   \langle ( \bar{A}_{k,x}^{\dagger} - \bar{A}_{k - 1,x})^{2}\rangle_{\text{eq}}
  & = \sum_{j,l = -\infty}^{\infty}
  \left(1 - \frac{\bar{\phi}_{\infty}}{\phi} \right)^{|j| + |l|}
  \langle U_{k + j,x} U_{k + l,x} \rangle_{\text{eq}} \\
  & = \sum_{j = -\infty}^{\infty}
  \left[ (1 + 2 \phi)
  \left(1 - \frac{\bar{\phi}_{\infty}}{\phi} \right)^{2 |j|}
  - 2 \phi \left(1 - \frac{\bar{\phi}_{\infty}}{\phi} \right)^{|j| + |j + 1|}
  \right] \\
  & = \frac{2 (1 + 2 \phi)}{1 - (1 - \bar{\phi}_{\infty} / \phi)^{2}}
  - (1 + 2 \phi)
  -\frac{ 4 \phi  (1 - \bar{\phi}_{\infty} / \phi)}{1 - (1 - \bar{\phi}_{\infty} / \phi)^{2}} \\
  & = 1 + 2 \bar{\phi}_{\infty}.
 \end{split} 
\end{equation}

Finally we have the following simple expression for the plateau modulus
of the equidistant slip-spring model:
\begin{equation}
 \label{equidistant_spring_plateau_modulus}
 \frac{G_{N,\text{ES}}^{(0)}}{G_{0}}
  \approx \frac{1}{1 + 2 \bar{\phi}_{\infty}} = \frac{1}{\sqrt{1 + 4 \phi}} .
\end{equation}
As the slip-spring size parameter $\phi$ increases, the plateau modulus decreases.
If the slip-spring size parameter is sufficiently small, a slip-spring
behaves as a slip-link. At the limit of $\phi \to 0$, the fluctuation
of the bond vanishes and the plateau modulus of the equidistant slip-link model is
recovered. (This limit would be interpreted as the pure reptation model.)
It would be fair to mention that Rubinstein and Panyukov\cite{Rubinstein-Panyukov-1997} obtained
a similar factor by a slightly different calculation.
From eq~\eqref{equidistant_spring_plateau_modulus},
the plateau moduli for the ideal and equidistant slip-springs are roughly
estimated as
\begin{equation}
 \label{ideal_repulsive_spring_plateau_modulus}
 \frac{G_{N,\text{IS}}^{(0)}}{G_{0}}
  \approx\frac{11/15}{\sqrt{1 + 4 \phi}} , \qquad
 \frac{G_{N,\text{RS}}^{(0)}}{G_{0}}
  \approx\frac{0.8214}{\sqrt{1 + 4 \phi}} .
\end{equation}
Eqs~\eqref{equidistant_spring_plateau_modulus} and \eqref{ideal_repulsive_spring_plateau_modulus}
are the main results of this work. The ratio of the plateau modulus to
$G_{0}$ depends both on the interaction between slip-links and the 
slip-spring size parameter $\phi$.

\section{Simulation}

Although we obtained the analytic expressions for the plateau modulus
in the previous section (eqs~\eqref{equidistant_spring_plateau_modulus} and \eqref{ideal_repulsive_spring_plateau_modulus}),
they are based on the single-subchain
approximation and the decoupling approximation. Under the single-subchain approximation, correlations
between different subchains are ignored and thus the results would not
be accurate. Also, the decoupling approximation is generally not so accurate.
To check the accuracy of our theory, in
this section we perform single-chain simulations without these approximations.
Then we can directly evaluate the plateau moduli with eqs \eqref{linear_response_formula_slip_link}
and \eqref{linear_response_formula_slip_spring_approx_local_equilibrium}.

As we mentioned, there are various similar but different slip-link
and slip-spring type models. Here we limit ourselves to simple models, and
do not directly compare elaborated simulation models. We employ the single-chain slip-link
and slip-spring models (without single-subchain approximations) as simulation models.
The comparison with the literature data by some elaborated simulation models are
shown in Section~\ref{comparison_with_simulation_data}.

Here we describe the simulation scheme.
First we generate a polymer chain with slip-linked
points by the equilibrium probability distribution. We assume that the average number
of slip-links (or slip-springs) $Z_{0}$ is sufficiently
large. Then we can use the segment number distribution for a single subchain
as a very accurate approximation. The algorithm to generate a
slip-linked or slip-springed chain is as follows.
\begin{enumerate}
 \item Set the segment index $s = 0$ and the subchain index $k = 1$.
 \item \label{chain_generation_segment_number}
       Generate $n_{k}$ from the equilibrium
       distribution of the segment number distribution
       (eq \eqref{segment_number_distribution_single_subchain}).
 \item If $s + n_{k} < Z_{0}$, increase $s$ and $k$ as $s \to s +
       n_{k}$ and $k \to k + 1$, and then go to the step \ref{chain_generation_segment_number}.
 \item Set $Z = k$ and $n_{Z} = Z_{0} - s$.
 \item Set $\bm{R}_{0} = 0$.
 \item For $k = 1,2,\dots,Z$, generate the slip-linked or slip-springed
       positions $\bm{R}_{k}$ from the Gaussian distribution.
 \item For slip-springed chains, generate the anchoring points $
       \bm{A}_{k}$ ($k = 0,1,\dots,Z$) from the
       Gaussian distribution.
\end{enumerate}

To obtain the locally
equilibrated state after $\tau_{e}$, we evolve the system by the
Monte Carlo method.
For the ideal and repulsive models, the segment numbers $\lbrace n_{k}
\rbrace$ are equilibrated. (In the equidistant slip-link
and slip-spring models, the segment number in a subchain is constant
and thus this exchange is not performed.)
We employ the Metropolis type
acceptance and rejection probabilities.
The equilibration scheme is as follows.
\begin{enumerate}
 \item Randomly choose two subchains indices $j$ and $k$.
 \item Randomly sample the number of segments $\Delta n$ to be exchanged, from
       the uniform distribution. $\Delta n$ is uniformly distributed in
       the range
       \begin{equation}
	- n_{j} \le \Delta n \le n_{k} ,
       \end{equation}
       to avoid $n_{j}$ or $n_{k}$ being negative.
 \item Accept the exchange by the following probability:
       \begin{equation}
	P(n_{j} + \Delta n, n_{k} - \Delta n | n_{j}, n_{k}) =
	 \min \left\lbrace 1,
	       e^{- \Delta\mathcal{F}(j,k;\Delta n)} \right\rbrace .
       \end{equation} 
       where the free energy difference $\Delta\mathcal{F}(j,k;\Delta n)$ is
       given as
\begin{align}
 & \Delta\mathcal{F}_{\text{I}}(j,k;\Delta n)
   \equiv
  \frac{\Delta n}{2} \bigg[ \frac{(\bm{R}_{k} - \bm{R}_{k - 1})^{2}}{n_{k} (n_{k}
 - \Delta n)}
  - \frac{(\bm{R}_{j} - \bm{R}_{j - 1})^{2}}{n_{j} (n_{j} +
	   \Delta n)} \bigg]
 + \frac{3}{2} \ln \frac{(n_{j} + \Delta n)
  (n_{k} - \Delta n)}{n_{j} n_{k}} , \\
 & \Delta\mathcal{F}_{\text{R}}(j,k;\Delta n)
   \equiv
 \frac{\Delta n}{2} \bigg[ \frac{(\bm{R}_{k} - \bm{R}_{k - 1})^{2}}{n_{k} (n_{k}
	   - \Delta n)}
  - \frac{(\bm{R}_{j} - \bm{R}_{j - 1})^{2}}{n_{j} (n_{j} +
	   \Delta n)} \bigg] ,
\end{align}
       for the ideal (``I'') and repulsive (``R'') models, respectively.
\end{enumerate}

For slip-spring models, the positions of slip-springed points $\lbrace
\bm{R}_{k} \rbrace$ are also equilibrated. We randomly select one bead
$j$ and then resample $\bm{R}_{j}$ from the local equilibrium
distribution. Because the subchains and slip-springs are linear springs,
the local equilibrium distribution of $\bm{R}_{j}$ becomes the
Gaussian. The resampling scheme is as follows.
\begin{enumerate}
 \item Randomly select the index $j$.
 \item Sample the Gaussian random number vector $\bm{w}$ from the standard normal
       distribution. The first and second statistical moments of
       $\bm{w}$ are given as
       $\langle \bm{w} \rangle = 0$ and $\langle \bm{w} \bm{w} \rangle = \bm{1}$.
 \item Generate the new position of the $j$-th point $\bm{R}_{j}$ as
\begin{equation}
 \bm{R}_{j} \to
  \begin{cases}
   \displaystyle \frac{n_{j}^{-1} \bm{R}_{j - 1}
   + n_{j + 1}^{-1} \bm{R}_{j + 1}
   + \phi^{-1} \bm{A}_{j}}{n_{j}^{-1} + n_{j + 1}^{-1} + \phi^{-1}}
   + (n_{j}^{-1} + n_{j + 1}^{-1} + \phi^{-1})^{-1/2} \bm{w} &
   (1 \le j \le Z) , \\
   \displaystyle \frac{
   n_{1}^{-1} \bm{R}_{1}
   + \phi^{-1} \bm{A}_{0}}{n_{1}^{-1} + \phi^{-1}}
   + (n_{1}^{-1} + \phi^{-1})^{-1/2} \bm{w} &
   (j = 0) , \\
   \displaystyle \frac{n_{Z}^{-1} \bm{R}_{Z - 1}
   + \phi^{-1} \bm{A}_{Z}}{n_{Z}^{-1} + \phi^{-1}}
   + (n_{Z}^{-1} + \phi^{-1})^{-1/2} \bm{w} &
   (j = Z) .
  \end{cases}
\end{equation}
\end{enumerate}

After sufficient Monte
Carlo trials (the total number of trials should be sufficiently large so that
the local equilibrium state is realized), we calculate the product of the
stress tensors appearing in eq
\eqref{linear_response_formula_slip_link_approx_local_equilibrium} (for
slip-link models) or in eq
\eqref{linear_response_formula_slip_spring_approx_local_equilibrium}
(for slip-spring models).
Finally, by taking
the average over different samples (different random number sequences),
we have the plateau modulus.

We use the Mersenne Twister random number generator\cite{Matsumoto-Nishimura-1998} to generate random
numbers. We also use the standard Box-Muller method\cite{Devroye-book} and the
Marsaglia-Tsang method\cite{Marsaglia-Tsang-2000} to generate random numbers which obey the
normal and gamma distributions, respectively.
In this work, we set the number of average
subchains as $Z_{0} = 100$, the total number of Monte Carlo
trials as $N_{\text{MC}} = 10^{3} Z_{0}$, and the number of samples
(chains) as $N_{\text{samples}} = 10^{5}$. (The simulation results are not sensitive
to these parameters, as far as they are sufficiently large.)

\begin{figure}[tb!]
 \centering
 {\includegraphics[width=0.95\linewidth,clip]{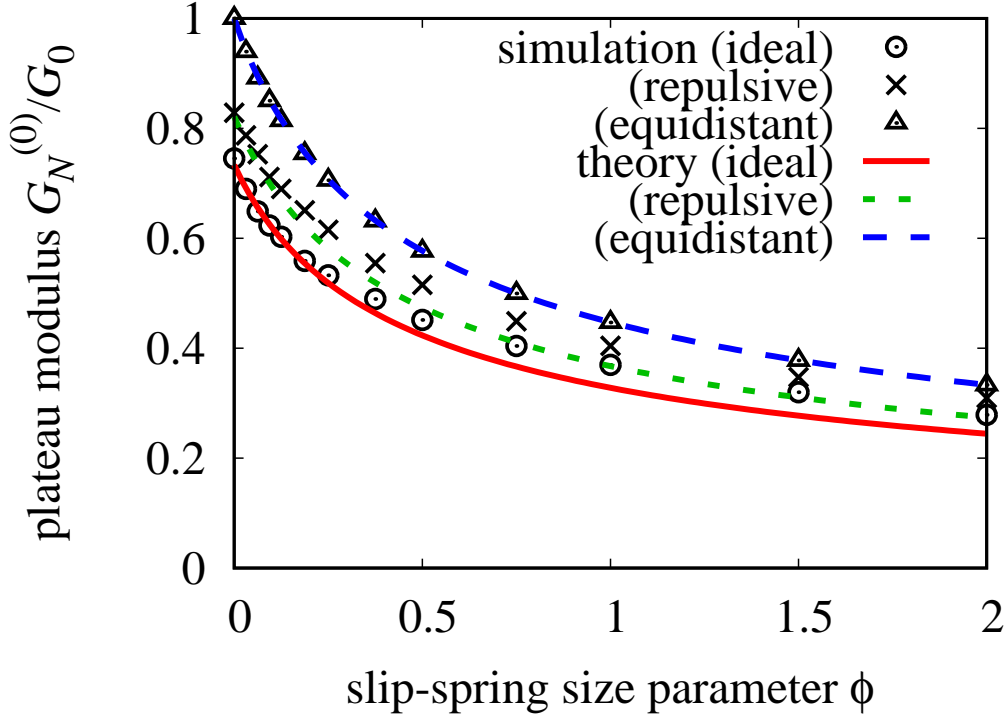}}
 \caption{Plateau moduli of several single-chain slip-link and slip-spring
models. Symbols show simulation data by the Monte Carlo method, and
curves show theoretical predictions (eqs~\eqref{equidistant_spring_plateau_modulus} and \eqref{ideal_repulsive_spring_plateau_modulus}).
The slip-link models correspond to the slip-spring models with
the slip-spring size parameter $\phi = N_{s} / N_{0} = 0$.}
 \label{plateau_modulus_simulation_and_theory}
\end{figure}

The Monte Carlo simulation data and theoretical predictions 
are shown in Figure~\ref{plateau_modulus_simulation_and_theory}. $\phi = 0$ corresponds
to the slip-link model. For the equidistant slip-spring model, the theoretical curve
coincides to the simulation data almost perfectly. For the repulsive and ideal slip-spring models,
the theoretical curves slightly deviate from the simulation data in the relatively large $\phi$ region.
For the slip-link models ($\phi = 0$), the agreement between the theoretical prediction
and the simulation data is very good. The deviation becomes larger as the slip-spring
size parameter $\phi$ increases, but even for relatively high $\phi$, the deviation
is not so large.
This result means that the single-subchain approximation is accurate for the
slip-link and slip-spring models. The deviation would be due to the coupling
of the fluctuations for the segment number and the bond vector.
(Judging from the simulation data, the decoupling approximation slightly
overestimates the fluctuations.)

\section{Discussions}
\label{discussions}

\subsection{Comparison with Simulation Data}
\label{comparison_with_simulation_data}

Here we compare our simulation results with literature data by some
slip-link and slip-spring models. Masubuchi et al\cite{Masubuchi-Ianniruberto-Greco-Marrucci-2003} performed
PCN simulations (the statistics of their model is similar
to the repulsive slip-link model), and reported that the plateau modulus is given by
$G_{N}^{(0)} = (4 / 5) G_{0}$ if slip-linked points
are fixed in space and only the segment numbers are allowed to be equilibrated.
(Although the PCN model is a multi-chain slip-link model, this result corresponds to the
single-chain model because slip-linked points are fixed and intrachain correlation is negligible.)
Nair and
Schieber\cite{Nair-Schieber-2006} reported that $N_{0}$ is well estimated from $G_{N}^{(0)}$ via
$G_{N}^{(0)} = (4 / 5) G_{0}$ for their slip-link model
(which corresponds to the ideal slip-link model in this work).
Although these results do not perfectly agree with our theory and Monte Carlo
simulation data, the decrease of the plateau modulus by the factor $4/5 =
0.8$ is close to our result ($G_{N}^{(0)} / G_{0} = 0.7333$ or $0.8214$).
(Strictly speaking, the plateau modulus at $t \approx \tau_{e}$ in the PCN model\cite{Masubuchi-Ianniruberto-Greco-Marrucci-2003}
seems to be slightly larger than $0.8$. This is consistent with our prediction.)
For the slip-spring model, the ideal
single-chain slip-spring simulations with $N_{0} = 4$ and $N_{s} = 0.5$ ($\phi =
0.125$) give roughly $G_{N}^{(0)} / G_{0} \approx
0.4$.\cite{Likhtman-2005,Uneyama-2011}
Our theory and the Monte Carlo simulation result give $G_{N}^{(0)} / G_{0}
 \approx 0.60$. This value is somewhat larger than the simulation result.
However, the estimate of the plateau modulus from the simulation data involves
some errors, and we consider the agreement is not that bad.
For the ideal slip-spring model, Likhtman\cite{Likhtman-2005}
examined the effects of $N_{0}$ and $N_{s}$ on the plateau modulus, and
reported some combinations of $N_{0}$ and $N_{s}$
which give almost the same plateau modulus; $(N_{0}, N_{s}) = (1, 4), (2, 1.75),$ and $(4, 0.5)$.
These parameter sets give the slip-spring
size parameters as $\phi = 4, 0.875,$ and $0.125$, respectively.
Our theory predicts the plateau moduli for
these parameter sets as $G_{N}^{(0)} / \rho_{0} k_{B} T \approx 0.18, 0.17,$
and $0.15$, respectively, and these values are reasonably close to each other.
(Note that here we have normalized $G_{N}^{(0)}$ by $\rho_{0} k_{B} T$ instead
of $G_{0} = \rho_{0} k_{B} T / N_{0}$. The value of $G_{N}^{(0)} / G_{0}$ strongly depends on $\phi$.)
Judging from these results, we consider that
our theory is reasonable to estimate the plateau moduli of various slip-link
and slip-spring models.

Even if our theory works well, we should note that the comparison shown above was
done for the single-chain models. In reality, the entanglement originates
from the non-crossability of polymer chains, and the validity of the
single-chain models is not guaranteed {\it a priori}. The
relations between the multi-chain slip-link and slip-spring models and the
corresponding single-chain models are not clear. Naively, we expect
that the fluctuations of slip-linked or slip-springed points in multi-chain models
are stronger than those of single-chain models.
We consider that such fluctuations are similar to those in cross-linked
networks. For the case of cross-linked networks, the shear modulus of a network can be
related to the fluctuations of cross-linked points\cite{Everaers-1999}.
For the ideal phantom $f$-functional random network, the shear modulus $G$
can be expressed as
\begin{equation}
 \label{rubber_elasticity_functionality_effect}
 G = (1 - 2 / f) \nu_{0} k_{B} T
\end{equation}
where $\nu_{0}$ is the average subchain number density.
Everaers\cite{Everaers-2012} analyzed series of network structures in molecular dynamics simulations
extracted by primitive path analyses, and reported that the plateau moduli can
be well reproduced by eq~\eqref{rubber_elasticity_functionality_effect} with $f = 4$.
If we assume that the similar expression holds for the multi-chain slip-link
models, the plateau modulus is reduced by the factor $(1 - 2 / f)$ from the
prediction for the single-chain slip-spring model.
At the single-chain level, this reduction would be
re-interpreted as the fluctuation of the slip-spring.
As we mentioned, we expect that slip-spring models can reproduce
some of the fluctuation effects. The degree of the fluctuation depends on
the slip-spring size parameter $\phi$ in the slip-spring model whereas it depends on the functionality $f$
in the phantom network model. Then we may simply employ the following relation:
$1 / \sqrt{1 + 4 \phi} \approx 1 - 2 / f$, which gives $\phi \approx (f - 1) / (f - 2)^{2}$.
For tri- and tetra-functional networks ($f = 3$ and $4$), we have
$\phi \approx 2$ and $3 / 4$, respectively.
With this interpretation, the PCN model
corresponds to the repulsive slip-spring model with $f = 4$, and thus our
theory predicts $G_{N}^{(0)} / G_{0} \approx 0.41$. This estimate seems to be
consistent with the PCN simulation data.\cite{Masubuchi-Ianniruberto-Greco-Marrucci-2003}
The multi-chain slip-spring (MCSS) model\cite{Uneyama-Masubuchi-2012}
corresponds to the ideal slip-spring model with $f = 3$, and our
theory predicts $G_{N}^{(0)} / G_{0} \approx 0.24$. This value seems to be
somewhat larger than the simulation data ($G_{N}^{(0)} / G_{0} \approx 0.17$) yet
still not that bad.

Recently, the detailed comparison among the molecular dynamics model,
the PCN model, and the MCSS model
was reported\cite{Masubuchi-Uneyama-2018}.
The relation between
the plateau moduli was shown to be $(G_{N,\text{PCN}}^{(0)} / G_{0}) / (G_{N,\text{MCSS}}^{(0)} / G_{0}) = 3.2$.
This value is also somewhat larger than the prediction by our theory, $0.41 / 0.24 = 1.7$.
For the MCSS model, our theory seems to overestimate the plateau modulus.
There are several possible reasons for this. In the MCSS model, two
ends of a slip-spring can be attached to the same polymer chain. Then,
for some conformations, slip-springs and some subchains
may not contribute to the stress. Such slip-springs and subchains decrease
the stress compared with a naive estimate. Also, the motion of slip-springs
is rather strongly coupled to the motion of polymer chains. The coupling
between different fluctuation mechanisms may enhance the fluctuations,
and the fluctuations in the MCSS model may be larger than our estimate.
(According to the simulation data with different effective fugacities,
the values of $G_{N,\text{MCSS}}^{(0)} / G_{0}$ are approximately constant and
thus independent of the effective fugacity\cite{Masubuchi-Doi-Uneyama-inpreparation}.
Thus the fluctuation mechanisms in the MCSS would not be sensitive to parameters in the MCSS model.)

The value of $G_{N}^{(0)} / G_{0}$ for each model can
be also estimated by fitting the
simulation data to experimental data\cite{Masubuchi-Ianniruberto-Greco-Marrucci-2008,Masubuchi-2018,Masubuchi-Uneyama-2019}.
In the case of the PCN model,
the fitting gives $G_{N}^{(0)} / G_{0} \approx 0.6$\cite{Masubuchi-Ianniruberto-Greco-Marrucci-2008}.
This value is larger than the estimate based only on the simulation data, and is also larger than our theoretical prediction.
In the case of the MCSS model, the fitting to experimental data gives
$G_{N}^{(0)} / G_{0} \approx 0.3$.
With these values we have $(G_{N,\text{PCN}}^{(0)} / G_{0}) / (G_{N,\text{MCSS}}^{(0)} / G_{0}) = 2$,
which seems to be close to the theoretical prediction.
However, it would be fair to mention that 
the fitting processes usually involve some errors and uncertainties, and the estimated values
depend on the details of the analysis and fitting methods.
We expect that our theoretical results work as reasonable estimates
for $G_{N}^{(0)} / G_{0}$ in various slip-link and slip-spring models,
within a certain error range.

Our result suggests that the average number of segments between
slip-links (or slip-springs), or other similar quantities which
characterizes the topological constraints, can {\em not} be calculated only
from the plateau modulus (unless the detailed fluctuation
mechanism of the model is fully known).
The plateau moduli of slip-link and slip-spring models are 
smaller than the modulus of the ideal rubber network with the average
segment number $N_{0}$,
unless the fluctuations are totally suppressed by such as the strong repulsive potential.
If we define the entanglement segment number
$N_{e}$ by eq~\eqref{segment_number_between_entanglements_definition},
$N_{e} /N_{0} = G_{0} /G_{N}^{(0)}$ and thus $N_{e}$ is larger than $N_{0}$.
As we showed, the ratio $G_{N}^{(0)} / G_{0}$ is a model-dependent quantity,
and therefore $N_{e} / N_{0}$ is also model-dependent.
Especially, in reality, there are no
slip-links nor slip-springs in entangled polymeric systems, and thus the
model-specific interpretation of the plateau modulus may lead
physically incorrect conclusions.
We should be careful to calculate or
estimate $N_{0}$ from experimental linear viscoelasticity data.

In recent years, the entanglement segment number and related physical quantities
are widely studied by the primitive path analyses\cite{Everaers-Sukumaran-Grest-Svaneborg-Sivasubramanian-Kremer-2004,Sukumaran-Grest-Kremer-Everaers-2005,Kroger-2005,Tzoumanekas-Theodorou-2006}.
Although there are 
several different primitive path analysis methods, the central idea is
common. The primitive paths can be extracted by virtually fixing the chain
ends and stretching chains tightly. Then, from the thus obtained primitive
paths, some statistical quantities such as the average contour length
and the statistical distribution of the topological constraints can be directly calculated.
We can estimate the entanglement segment number by the primitive path, $N_{\text{PP}}$,
from these quantities.
The primitive path analyses have been applied for various molecular
models such as the atomistic and coarse-grained Kremer-Grest molecular dynamics models.
The entanglement segment number from the primitive path, $N_{\text{PP}}$,
seems to be smaller than that calculated from the plateau modulus, $N_{e}$.
Everaers\cite{Everaers-2012} collected the literature data and calculated
$N_{\text{PP}} / N_{e}$ for various molecular models with various primitive path analysis methods.
According to his data, $N_{\text{PP}} / N_{e}$ seems to be roughly about $0.5$.
However, the reported values of $N_{\text{PP}} / N_{e}$ are not exactly the same value but
seem to depend on the details of the analysis method and the system.
Therefore, we consider that $N_{\text{PP}}$ is rather similar to
$N_{0}$ in slip-link and slip-spring models.
The difference among different primitive path extraction methods may result in different
statistical properties. We expect that, as the case of the slip-link and slip-spring
models, the ratio $N_{\text{PP}} / N_{e}$ would reflect something like fluctuations.
The ratio $N_{\text{PP}} / N_{e}$
would be an important quantity when we compare the coarse-grained models and
molecular dynamics simulations.
Steenbakkers et al\cite{Steenbakkers-Tzoumanekas-Li-Liu-Kroger-Schieber-2014}
proposed a method to map the primitive path to the single-chain slip-link model.
Recent work by Becerra et al\cite{Becerra-Cordoba-Katzarova-Andreev-Venerus-Schieber-2020}
showed that an atomistic molecular model for polyethylene oxide (PEO) can be quantitatively mapped onto
some single-chain slip-link models. Although they showed good agreement between experimental and simulation relaxation modulus data,
whether their method works well for other systems such as the slip-spring model
is not clear.
To realize such a mapping, we should understand statistical properties of target models in detail.
Our results may be utilized to develop similar mapping method for various slip-link and
slip-spring models.

\subsection{Relaxation by Longitudinal Motion}

In our theory, the effect of the longitudinal motion\cite{Doi-Edwards-book} is not
taken into account. This is because at the time scale of $t \approx \tau_{e}$,
the longitudinal motion cannot occur.
(Sometimes the relaxation motion of the longitudinal motion is also referred as
the contour length fluctuation (CLF). Here, to make the relaxation mechanism clear, we
use the term longitudinal motion.)
However, the reduction of the stress by the longitudinal motion is widely employed
in the analyses for the plateau modulus. According to the standard tube theory,
the plateau modulus is modulated from the classical formula and given as\cite{Doi-Edwards-book}
\begin{equation}
 \label{plateau_modulus_with_contour_length_fluctuation}
 G_{N}^{(0)} = \frac{4}{5} \frac{\rho_{0} k_{B} T}{N_{e}'} .
\end{equation}
Here, $N_{e}'$ represents the characteristic segment number which characterizes the tube segment size.
(We note that eq~\eqref{plateau_modulus_with_contour_length_fluctuation} was calculated essentially in the same way as
eq~\eqref{plateau_modulus_and_relaxation_modulus}.\cite{Doi-1983})
If the longitudinal motion is absent, the plateau modulus becomes $G_{N}^{(0)} = \rho_{0} k_{B} T / N_{e}'$ (the characteristic segment
number $N_{e}'$ is assumed to be the same as that in eq~\eqref{plateau_modulus_with_contour_length_fluctuation}).
Thus the longitudinal motion reduces the plateau modulus by the factor $4/5$, in a similar way to the fluctuations of
segment numbers and slip-springed points in our model.
Likhtman and McLeish\cite{Likhtman-McLeish-2002} analyzed the
longitudinal modes of a polymer chain in a tube in detail, and showed that the
longitudinal motion does not give the decrease of the plateau modulus at
the time scale $t \lesssim \tau_{e}$.
This is physically natural because the characteristic time scale for the relaxation of the longitudinal
modes is $\tau_{R}$. (Their analysis is based
on the tube model and thus it may not be fully applicable to the slip-link and
slip-spring models. However, the characteristic time scale of the longitudinal
modes is common for other models.)
Therefore, from the view point of the time scale, the theoretical
validity of eq~\eqref{plateau_modulus_with_contour_length_fluctuation} is
somewhat questionable.

Our analysis showed that the decrease of the plateau modulus is caused
by the fluctuations (or relaxation) of segment numbers and
slip-springed points, which occur at the short time scale. Thus our result is not inconsistent with the Likhtman and
McLeish's analysis. (Roughly speaking, the tube model can be interpreted
as the equidistant slip-link model in which the plateau modulus is simply given as $G_{0}$.)
As we mentioned, our theory predicts the numerical
factor $0.7333$ or $0.8214$ at the time scale of $t \approx \tau_{e}$.
These factors are close to $4/5$.
In addition, if we tune the repulsive interaction between slip-links precisely,
we may be able to perfectly reproduce the factor $4/5$.

Therefore, it would be dangerous to blindly employ eq~\eqref{plateau_modulus_with_contour_length_fluctuation}.
The segment number between entanglements is sometimes estimated as $N_{e}'$
by utilizing eq~\eqref{plateau_modulus_with_contour_length_fluctuation} (so-called the Graessley type definition)\cite{Fetters-Lohse-Graessley-1999,Larson-Sridhar-Leal-McKinley-Likhtman-McLeish-2003}:
\begin{equation}
 \label{segment_number_between_entanglements_definition_with_clf}
 N_{e}' = \frac{4}{5} \frac{\rho_{0} k_{B} T}{G_{N}^{(0)}} = \frac{4}{5} N_{e},
\end{equation}
However, according to the discussions above, this seems not to be sound.
The definition of the segment number between entanglements should not depend on a specific theoretical interpretation\cite{Osaki-2002}.
If we employ eq~\eqref{segment_number_between_entanglements_definition_with_clf}
as the definition of the segment number between entanglement
instead of $N_{e}$ by eq~\eqref{segment_number_between_entanglements_definition},
the physical interpretation of the segment number between entanglements
becomes rather complicated.
(We will need to introduce an extra numerical factor $5/4$ to relate $N_{0}$ to this segment number between entanglement $N_{e}'$.)
Based on the results shown above, we conclude that the classical and traditional definition by eq~\eqref{segment_number_between_entanglements_definition}
should be employed for the segment number between entanglements.
The relaxation mechanisms which have longer time scales than $\tau_{e}$ should not
be considered when we study the properties of the plateau modulus.

\section{Conclusions}

We theoretically calculated the plateau moduli of the single-chain slip-link and slip-spring
models. We employed several approximations such as the single-subchain
approximation and the decoupling approximation, to analytically calculate
the plateau moduli.
We showed that the plateau modulus depends on various
factors such as the interaction between slip-links and the spring constant
of a slip-spring.
The plateau modulus increases as the repulsive interaction
between slip-links or slip-springs increases. Also, the plateau modulus
decreases as the slip-spring size parameter increases.
We obtained quantitative predictions which relate parameters in
slip-link and slip-spring models to the plateau modulus.

We compared our theoretical results
(eqs~\eqref{equidistant_spring_plateau_modulus} and \eqref{ideal_repulsive_spring_plateau_modulus})
with the Monte Carlo simulation results
and the simulation data in the literature. The agreement between theoretical results
and Monte Carlo simulation results is good. Our theory slightly underestimates
the plateau moduli when the slip-spring size parameter is large.
From the comparison of the theoretical results and the literature data,
we found that our theory can roughly explain the simulation data (although
the agreement is not perfect). Conventionally, the reduction of the plateau
modulus is mainly interpreted as the effect of the longitudinal motion. Our theory gives a new
and physically reasonable interpretation for the reduction of the
plateau modulus.

From the results of this work, we conclude that the segment number between 
entanglements $N_{e}$ and the segment number between two slip-links or slip-springs
$N_{0}$ are generally different and the relation between them is model-dependent. We consider that this work justifies to
use $N_{0}$ and $N_{e}$ (or $G_{N}^{(0)}$) as independent fitting parameters
for slip-link and slip-spring models. The ratio $N_{0} / N_{e}$ reflects
short time fluctuation mechanisms, and thus it is strongly
model-dependent. This ratio would be useful when we compare several
different entanglement models and check the consistency among them.
It would be also useful to map one simulation model to
another simulation model and perform multi-scale simulations.

\section*{Acknowledgment}

This work was supported by
Grant-in-Aid (KAKENHI) for Scientific Research Grant B Number JP19H01861,
Grant-in-Aid (KAKENHI) for Scientific Research on Innovative Areas 
``Discrete Geometric Analysis for Materials Design'' Grant Number JP20H04636
(from Ministry of Education, Culture, Sports, Science, and Technology),
JST-PRESTO Grant Number JPMJPR1992,
and JST-CREST Grant Number JPMJCR1992 (from Japan Science and Technology Agency).

\appendix

\section*{Appendix}

\section{Simple Derivation of Equilibrium Probability Distributions for Single
 Subchain}
\label{simple_derivation_of_probability_distributions_for_single_subchain}

In this appendix, we derive equilibrium probability distribution
functions for a single subchain in the single-chain slip-link models.
Following the idea by Greco\cite{Greco-2008}, we introduce the
effective chemical potential to control the number of segments in a subchain.
Then we can directly calculate the statistical properties of a single subchain.
We employ the same dimensionless units in the main text.
The joint probability distribution of the segment number $n$ and the bond
vector $\bm{Q}$ for a subchain is given as:
\begin{equation}
 \label{probability_distribution_single_subchain_grand_canonical}
 P_{\text{eq}}(\bm{Q},n) \propto
  e^{- \mathcal{F}_{\text{single}}(\bm{Q},n) - \mu n} ,
\end{equation}
where $\mathcal{F}_{\text{single}}(\bm{Q},n)$ is the free energy for a single subchain
(including the effective interaction between slip-links), $\mu$ is the effective chemical potential which is determined to
satisfy the condition $\langle n \rangle_{\text{eq}} = 1$. The single
subchain free energy is given as follows:
\begin{equation}
 \label{free_energy_single_subchain}
 \mathcal{F}_{\text{single}}(\bm{Q},n) = \frac{\bm{Q}^{2}}{2 n}
  + \left(\frac{3}{2} - \alpha\right) \ln n .
\end{equation}
Here the parameter $\alpha$ represents the strength
of the repulsive interaction between slip-links\cite{Uneyama-Masubuchi-2011}. The ideal and repulsive
slip-link models correspond to $\alpha = 0$ and $3/2$, respectively. The
equidistant slip-link model corresponds to the limit of $\alpha \to \infty$.
The normalization constant for the probability given by eq
\eqref{probability_distribution_single_subchain_grand_canonical} (the grand
partition function) is
calculated straightforwardly:
\begin{equation}
 \label{probability_distribution_single_subchain_grand_canonical_normalization_constant}
   \int d\bm{Q} dn \, e^{- \mathcal{F}_{\text{single}}(\bm{Q},n) - \mu n}
   = (2 \pi)^{3/2} \mu^{- \alpha - 1} \Gamma( \alpha + 1) ,
\end{equation}
where $\Gamma(x)$ is the gamma function\cite{Abramowitz-Stegun-book,NIST-handbook}.
From eqs
\eqref{probability_distribution_single_subchain_grand_canonical} and
\eqref{probability_distribution_single_subchain_grand_canonical_normalization_constant},
the equilibrium distribution function $P_{\text{eq,L}}(\bm{Q},n)$ is rewritten as the following explicit form:
\begin{equation}
 \label{probability_distribution_single_subchain_grand_canonical_modified}
   P_{\text{eq},\text{L}}(\bm{Q},n)
    = \frac{\mu^{\alpha + 1}}{(2 \pi)^{3/2} \Gamma( \alpha + 1)}
    n^{\alpha - 3 / 2}   e^{- \bm{Q}^{2} / 2 n - \mu n} .
\end{equation}

The single-subchain level statistical properties can be calculated from
eq~\eqref{probability_distribution_single_subchain_grand_canonical_modified}.
The segment number distribution function is obtained by integrating
eq~\eqref{probability_distribution_single_subchain_grand_canonical_modified}
over $\bm{Q}$:
\begin{equation}
 \label{segment_number_distribution_single_subchain_grand_canonical}
  P_{\text{eq},\text{L}}(n)
   = \int d\bm{Q} \, P_{\text{eq},\text{L}}(\bm{Q},n)
   = \frac{\mu^{\alpha + 1} }{\Gamma( \alpha + 1)} n^{\alpha}  e^{- \mu n} .
\end{equation}
Thus the equilibrium average of the segment number becomes
\begin{equation}
 \label{average_segment_number_single_subchain_grand_canonical}
  \langle n \rangle_{\text{eq}} = \int dn \, n P_{\text{eq},\text{L}}(n)
  = \frac{\alpha + 1}{\mu} .
\end{equation}
From the condition $\langle n \rangle_{\text{eq}} = 1$, we have the simple relation
between $\mu$ and $\alpha$, as $\mu = \alpha + 1$.
Finally the segment number distribution
\eqref{segment_number_distribution_single_subchain_grand_canonical}
is rewritten as
\begin{equation}
 \label{segment_number_distribution_single_subchain_grand_canonical_modified}
  P_{\text{eq},\text{L}}(n)
   = \frac{(\alpha + 1)^{\alpha + 1} }{\Gamma( \alpha + 1)} n^{\alpha} e^{- (\alpha + 1) n} .
\end{equation}
Eq~\eqref{segment_number_distribution_single_subchain_grand_canonical_modified}
gives eq~\eqref{segment_number_distribution_single_subchain} in the main text.
(At the limit of $\alpha \to \infty$,
eq \eqref{segment_number_distribution_single_subchain_grand_canonical_modified}
approaches to the delta function.)

The bond vector distribution function is obtained by integrating eq
\eqref{probability_distribution_single_subchain_grand_canonical_modified}
over $n$. For $\alpha = 0$, by using the variable transform $u = 1 / \sqrt{n}$, we have
\begin{equation}
\begin{split}
 \label{bond_vector_distribution_single_subchain_grand_canonical_ideal}
 P_{\text{eq},\text{IL}}(\bm{Q})
 & = \frac{2}{(2 \pi)^{3/2}} \int_{0}^{\infty} du \, 
 e^{- \bm{Q}^{2}  u^{2} / 2 - u^{-2}} \\
 & =
 \frac{1}{4 \pi |\bm{Q}|}
 \bigg[ e^{\sqrt{2} |\bm{Q}|} \operatorname{erf}\bigg( \frac{|\bm{Q}| u}{\sqrt{2}}
 + \frac{1}{u} \bigg)
 + e^{- \sqrt{2} |\bm{Q}|} \operatorname{erf}\bigg( \frac{|\bm{Q}| u}{\sqrt{2}}
 - \frac{1}{u} \bigg) \bigg]_{0}^{\infty}
 \\
 & = \frac{1}{2 \pi |\bm{Q}|}
 e^{- \sqrt{2} |\bm{Q}|} ,
\end{split}
\end{equation}
where $\operatorname{erf}(x)$ is the error function and we have used the integral formula for the error function\cite{Abramowitz-Stegun-book,NIST-handbook}.
For $\alpha = 3 / 2$, by introducing the variable transform $s = \ln
(\sqrt{5} n  / |\bm{Q}|)$, we have
\begin{equation}
 \label{bond_vector_distribution_single_subchain_grand_canonical_repulsive}
  \begin{split}
   P_{\text{eq},\text{RL}}(\bm{Q})
   & = \frac{25}{12 \pi^{2}} |\bm{Q}|  \int_{-\infty}^{\infty} ds \, e^{s -
   \sqrt{5} |\bm{Q}| \cosh s } \\
   & = \frac{25}{6 \pi^{2}} |\bm{Q}| K_{1}(\sqrt{5} |\bm{Q}|) .
  \end{split}
\end{equation}
Here we have utilized the integral representation of the modified Bessel
function\cite{Lebedev-book}.
For the equidistant slip-link model (at the limit of $\alpha \to
\infty$), the $n$-dependent term in the integrand approaches to the delta function.
Thus we simply have
\begin{equation}
 \label{bond_vector_distribution_single_subchain_grand_canonical_equidistant}
 P_{\text{eq},\text{EL}}(\bm{Q}) = \frac{1}{(2 \pi)^{3/2}} e^{- \bm{Q}^{2} / 2} .
\end{equation}
Eqs
\eqref{bond_vector_distribution_single_subchain_grand_canonical_ideal}-\eqref{bond_vector_distribution_single_subchain_grand_canonical_equidistant}
give eq \eqref{bond_vector_distribution_single_subchain} in the main text.

\section{Characteristic Modulus}
\label{characteristic_modulus}

At $t = 0$, the shear relaxation modulus $G(t = 0)$ coincides to the
characteristic modulus $G_{0}$.
This property is common for all the slip-link and slip-spring models
examined in this work. 
Because $G(t = 0)$ is expressed as the equal-time correlation, it can be
straightforwardly evaluated.
For the slip-link models, we have
\begin{equation}
 \begin{split}
  G_{\text{L}}(t = 0)
  & = \nu_{0}
  \langle \hat{\sigma}_{xy}^{2} \rangle_{\text{eq}} \\
  & = \nu_{0} \sum_{Z = 1}^{\infty} \int
  d\lbrace n_{k} \rbrace \bigg[ \int d\lbrace \bm{R}_{k} \rbrace \, 
  \sum_{k = 1}^{Z} \frac{Q_{k,x}^{2} Q_{k,y}^{2}}{n_{k}^{2}} \prod_{k = 1}^{Z} P_{\text{eq},\text{L}}(
  \bm{R}_{k} | n_{k}) \bigg]  P_{\text{eq},\text{L}}(\lbrace n_{k} \rbrace, Z) \\
  & = \nu_{0} \sum_{Z = 1}^{\infty} Z 
  \int d\lbrace n_{k} \rbrace \,
  P_{\text{eq},\text{L}}(\lbrace n_{k} \rbrace, Z) = G_{0} .
 \end{split}
\end{equation}
On the other hand, for the slip-spring models, we have
\begin{equation}
 \begin{split}
  G_{\text{S}}(t = 0)
  & = \nu_{0} \langle \hat{\sigma}_{xy}^{2} \rangle_{\text{eq}}
  + \nu_{0} \langle \hat{\sigma}_{xy} \hat{\sigma}_{xy}^{(v)} \rangle_{\text{eq}}
  \\
  & = G_{0} + \nu_{0} \sum_{Z = 1}^{\infty} \int d\lbrace n_{k} \rbrace \, \bigg[ 
  \int d\lbrace \bm{R}_{k} \rbrace \, \sum_{k = 1}^{Z} \frac{Q_{k,x} Q_{k,y}}{n_{k}}   \prod_{k = 1}^{Z}
  P_{\text{eq},\text{S}}(\bm{R}_{k} | n_{k}) \bigg] \bigg[ \int d\lbrace \bm{R}_{k} \rbrace \\
  & \qquad \times 
    \sum_{k = 0}^{Z} \frac{(R_{k,x} - A_{k,x})  (R_{k,y} - A_{k,y})}{\phi} 
 P_{\text{eq},\text{S}}(\lbrace \bm{R}_{k} \rbrace,\lbrace \bm{A}_{k} \rbrace | \lbrace n_{k} \rbrace, Z ) \bigg]
  P_{\text{eq},\text{S}}(\lbrace n_{k} \rbrace, Z)
  \\
  & = G_{0} .
 \end{split}
\end{equation}
Therefore, for all models examined in this work, $G(t = 0)
= G_{0}$.
Of course, in reality, $G(t = 0)$ does not coincide to $G_{0}$ because
there are fast relaxation modes such as the subchain-scale Rouse modes and
segment (glassy) modes.
The characteristic modulus calculated here ignores contributions of these fast modes, and
only the entanglement modes are considered.
The slip-link and slip-spring models examined in this
work are coarse-grained models, and the fine scale
relaxations (or fluctuations) are not explicitly taken into account.
If we integrate some fast relaxation modes into the model, $G_{0}$ will
be changed.
(In this sense, one may interpret the characteristic modulus $G_{0}$ as
a rather model dependent quantity.)

\section{Single-Subchain Statistics in Equidistant Slip-Spring Model}
\label{single_subchain_statistics_in_equidistant_slip_spring_model}

In this appendix, we show the detailed calculations for the
single-subchain properties in the equidistant slip-spring model.
In this model, the number of segments in a subchain is constant, and
thus we do not need to consider the degrees of freedom for the segment numbers. (For sufficiently long chains, the effects of the chain ends
become negligibly small. Thus we assume that the segment numbers in
subchains at chain ends are also constant.)
The equilibrium distribution function can
be expressed as
\begin{equation}
 \label{equilibrium_distribution_full_chain_equidistant_slip_spring}
 P_{\text{eq},\text{ES}}(\lbrace \bm{R}_{k} \rbrace,\lbrace \bm{A}_{k} \rbrace)
  \propto \exp\bigg[
  - \frac{1}{2} \sum_{k = 1}^{Z} (\bm{R}_{k} - \bm{R}_{k - 1})^{2}
  - \frac{1}{2 \phi} \sum_{k = 0}^{Z} (\bm{R}_{k} - \bm{A}_{k})^{2}
  \bigg] .
\end{equation}
In principle,
we can obtain the probability distribution for a single subchain by integrating
eq~\eqref{equilibrium_distribution_full_chain_equidistant_slip_spring}
over $\lbrace \bm{R}_{k} \rbrace$ except the target subchain.
However, since all the anchoring points are (indirectly) connected (by the polymer
chain), such an integration is not that simple.
The difficulty mainly arises from the coupling between $\lbrace \bm{R}_{k} \rbrace$
and $\lbrace \bm{A}_{k} \rbrace$. The integral over $\lbrace \bm{R}_{k} \rbrace$
is a Gaussian integral over multiple variables, and we need to calculate
the covariance matrix and its inverse. Unfortunately, 
the covariance matrix is not in a simple form in the current case.

Here we calculate the integral in an iterative manner, by starting from the chain end.
The integral over $\bm{R}_{0}$ can be calculated straightforwardly:
\begin{equation}
  \label{equilibrium_distribution_full_chain_equidistant_slip_spring_integral_r0}
\begin{split}
 & \int d\bm{R}_{0} \, P_{\text{eq},\text{ES}}(\lbrace \bm{R}_{k} \rbrace,\lbrace \bm{A}_{k} \rbrace) \\
 & \propto \exp\bigg[
  - \frac{1}{2 (1 + \phi)} (\bm{R}_{1} - \bm{A}_{0})^{2}
  - \frac{1}{2} \sum_{k = 2}^{Z} (\bm{R}_{k} - \bm{R}_{k - 1})^{2}
  - \frac{1}{2 \phi} \sum_{k = 1}^{Z} (\bm{R}_{k} - \bm{A}_{k})^{2}
  \bigg] \\ 
 & \propto \exp\bigg[
  - \frac{1}{2 \bar{\phi}_{1}}
 (\bm{R}_{1} - \bar{\bm{A}}_{1})^{2}
 - \frac{1}{2 \phi} (\bm{A}_{0}^{2} + \bm{A}_{1}^{2})
 + \frac{1}{2 \bar{\phi}_{1}} \bar{\bm{A}}_{1}^{2} \\
 & \qquad - \frac{1}{2} \sum_{k = 2}^{Z} (\bm{R}_{k} - \bm{R}_{k - 1})^{2}
  - \frac{1}{2 \phi} \sum_{k = 2}^{Z} (\bm{R}_{k} - \bm{A}_{k})^{2}
  \bigg] ,
\end{split}
\end{equation}
with
\begin{align}
 \label{effective_slip_spring_size_r1}
 \frac{1}{\bar{\phi}_{1}} & \equiv \frac{1}{\phi} + \frac{1}{1 + \phi} , \\
 \label{effective_anchoring_point_r1}
 \bar{\bm{A}}_{1} & \equiv \frac{\bar{\phi}_{1}}{\phi} \bm{A}_{1} + 
 \left( \frac{\bar{\phi}_{1}}{\phi} - 1 \right) \bm{A}_{0} .
\end{align}
We may interpret eq~\eqref{equilibrium_distribution_full_chain_equidistant_slip_spring_integral_r0}
as the change of the anchoring point and the slip-spring size for $\bm{R}_{1}$.
Eqs~\eqref{effective_slip_spring_size_r1} and \eqref{effective_anchoring_point_r1}
give the effective anchoring position and the slip-spring size.
Then we can integrate eq~\eqref{equilibrium_distribution_full_chain_equidistant_slip_spring_integral_r0}
over $\bm{R}_{2}$ in the same manner.
In eq~\eqref{equilibrium_distribution_full_chain_equidistant_slip_spring_integral_r0},
only the terms which contain $\bm{R}_{0}$ and $\bm{R}_{1}$ are changed and
other therms are unchanged. The terms which do not depend on $\bm{R}_{2}$
never affect the integral over $\bm{R}_{2}$.
Therefore, we can safely ignore most of terms when we calculate the integral.
If we have performed the multiple integrals over $\bm{R}_{0},\bm{R}_{1},\dots,\bm{R}_{j - 1}$,
we should have
\begin{equation}
  \label{equilibrium_distribution_full_chain_equidistant_slip_spring_integral_rj1}
\begin{split}
 & \int d\bm{R}_{0}d\bm{R}_{1} \dots d\bm{R}_{j - 1} \, P_{\text{eq},\text{ES}}(\lbrace \bm{R}_{k} \rbrace,\lbrace \bm{A}_{k} \rbrace) \\
 & \propto \exp\bigg[
  - \frac{1}{2 \bar{\phi}_{j}}
 (\bm{R}_{j} - \bar{\bm{A}}_{j})^{2} -
 \frac{1}{2}(\bm{R}_{j} - \bm{R}_{j + 1})^{2}
  - \frac{1}{2 \phi}
 (\bm{R}_{j + 1} - \bar{\bm{A}}_{j + 1})^{2} -
 \dots
  \bigg] , 
\end{split}
\end{equation}
and the multiple integrals over $\bm{R}_{0},\bm{R}_{1},\dots,\bm{R}_{j}$ become
\begin{equation}
  \label{equilibrium_distribution_full_chain_equidistant_slip_spring_integral_rj}
\begin{split}
 & \int d\bm{R}_{0}d\bm{R}_{1} \dots d\bm{R}_{j} \, P_{\text{eq},\text{ES}}(\lbrace \bm{R}_{k} \rbrace,\lbrace \bm{A}_{k} \rbrace) \\
 & \propto \int d\bm{R}_{j} \, \exp\bigg[
  - \frac{1}{2 (1 + \bar{\phi}_{j})}
 (\bm{R}_{j + 1} - \bar{\bm{A}}_{j})^{2} 
  - \frac{1}{2 \phi}
 (\bm{R}_{j + 1} - \bar{\bm{A}}_{j + 1})^{2} -
 \dots
  \bigg] \\
 & \propto \exp\bigg[
  - \frac{1}{2}
 \left(\frac{1}{\phi} + \frac{1}{1 + \bar{\phi}_{j}}\right)
 \left[\bm{R}_{j + 1} -
 \frac{\bm{A}_{j + 1} / \phi + \bar{\bm{A}}_{j} / (1 + \bar{\phi}_{j})}{1 / \phi + 1 / (1 + \bar{\phi}_{j})} \right]^{2} -
 \dots
  \bigg] .
\end{split}
\end{equation}
From eq~\eqref{equilibrium_distribution_full_chain_equidistant_slip_spring_integral_rj},
we find the following recurrence relations for the effective anchoring point
and the slip-spring size:
\begin{align}
 \label{effective_slip_spring_size_recurrence}
 \frac{1}{\bar{\phi}_{j + 1}} & = \frac{1}{\phi} + \frac{1}{1 + \bar{\phi}_{j}}, \\
 \label{effective_anchoring_point_recurrence}
 \bar{\bm{A}}_{j + 1} & = \frac{\bm{A}_{j + 1} / \phi + \bar{\bm{A}}_{j} / (1 + \bar{\phi}_{j})}{1 / \phi + 1 / (1 + \bar{\phi}_{j})}
 = \frac{\bar{\phi}_{j + 1}}{\phi} \bm{A}_{j + 1}
 + \left(1 - \frac{\bar{\phi}_{j + 1}}{\phi}\right) \bar{\bm{A}}_{j}.
\end{align}
The ``initial'' conditions are $\phi_{0} = \phi$ and $\bar{\bm{A}}_{0} = \bm{A}_{0}$.
The integral can be performed in the opposite direction, from another chain end $(j = Z)$. In this case, we have
the effective slip-spring size $\bar{\phi}_{j}^{\dagger}$ and
the effective anchoring point $\bar{\bm{A}}^{\dagger}_{j}$ which satisfy
\begin{align}
 \label{effective_slip_spring_size_dagger_recurrence}
 \frac{1}{\bar{\phi}_{j - 1}^{\dagger}} & = \frac{1}{\phi} + \frac{1}{1 + \bar{\phi}_{j}^{\dagger}}, \\
 \label{effective_anchoring_point_dagger_recurrence}
 \bar{\bm{A}}_{j - 1}^{\dagger} & = 
  \frac{\bar{\phi}_{j - 1}^{\dagger}}{\phi} \bm{A}_{j - 1}
 + \left(1 - \frac{\bar{\phi}_{j - 1}^{\dagger}}{\phi}\right) \bar{\bm{A}}_{j}^{\dagger},
\end{align}
with the ``initial'' conditions $\bar{\phi}^{\dagger}_{Z} = \phi$ and $\bar{\bm{A}}_{Z}^{\dagger} = \bm{A}_{Z}$.

The recurrence relations \eqref{effective_slip_spring_size_recurrence}-\eqref{effective_anchoring_point_dagger_recurrence}
are not easy to solve analytically.
Thus we seek approximate solutions, instead of the exact solutions.
For a sufficiently long chain, the effective slip-spring size approximately becomes
constant, $\bar{\phi}_{j} \approx \bar{\phi}_{\infty}$, where $\bar{\phi}_{\infty}$ is defined via
\begin{equation}
 \label{effective_slip_spring_size_infinity}
 \frac{1}{\bar{\phi}_{\infty}}  = \frac{1}{\phi} + \frac{1}{1 + \bar{\phi}_{\infty}} .
\end{equation}
The solution of eq~\eqref{effective_slip_spring_size_infinity} is
\begin{equation}
 \bar{\phi}_{\infty} = \frac{-1 + \sqrt{1 + 4 \phi}}{2} .
\end{equation}
The recurrence relation \eqref{effective_anchoring_point_recurrence} can be
also approximated as
\begin{equation}
 \label{effective_anchoring_point_recurrence_approx}
 \bar{\bm{A}}_{j + 1} 
 \approx \frac{\bar{\phi}_{\infty}}{\phi} \bm{A}_{j + 1}
 + \left(1 - \frac{\bar{\phi}_{\infty}}{\phi}\right) \bar{\bm{A}}_{j},
\end{equation}
and eq~\eqref{effective_anchoring_point_recurrence_approx} can be easily
solved:
\begin{equation}
 \label{effective_anchoring_point_approx_solution}
 \bar{\bm{A}}_{j} \approx \frac{\bar{\phi}_{\infty}}{\phi} \sum_{k = 0}^{\infty}
  \left(1 - \frac{\bar{\phi}_{\infty}}{\phi} \right)^{j} \bm{A}_{j - k} .
\end{equation}
In a similar way, we have the approximate solutions for $\bar{\phi}_{j}^{\dagger}$
and $\bar{\bm{A}}_{j}^{\dagger}$: $\bar{\phi}_{j}^{\dagger} \approx \bar{\phi}_{\infty}$ and
\begin{equation}
 \label{effective_anchoring_point_dagger_approx_solution}
 \bar{\bm{A}}_{j}^{\dagger} \approx \frac{\bar{\phi}_{\infty}}{\phi} \sum_{k = 0}^{\infty}
  \left(1 - \frac{\bar{\phi}_{\infty}}{\phi} \right)^{j} \bm{A}_{j + k} .
\end{equation}
Under these approximations,
finally, the probability distribution for a single subchain is given as
\begin{equation}
 P_{\text{eq},\text{ES}}(\bm{R}_{k - 1},\bm{R}_{k},\lbrace \bm{A}_{k} \rbrace)
  \propto  \exp
  \left[ - \frac{(\bm{R}_{k} - \bm{R}_{k - 1})^{2}}{2} 
   - \frac{(\bm{R}_{k - 1} - \bar{\bm{A}}_{k - 1})^{2}}{2 \bar{\phi}_{\infty}} 
   - \frac{(\bm{R}_{k} - \bar{\bm{A}}^{\dagger}_{k})^{2}}{2 \bar{\phi}_{\infty}}
   + \dots
  \right] ,
\end{equation}
and the conditional probability distribution for $\bm{R}_{k - 1}$ and $\bm{R}_{k}$ becomes
\begin{equation}
\begin{split}
  P_{\text{eq},\text{ES}}(\bm{R}_{k - 1},\bm{R}_{k} | \lbrace \bm{A}_{k} \rbrace)
 & = \frac{(1 + 2 \bar{\phi}_{\infty})^{3/2}}{(2 \pi \bar{\phi}_{\infty})^{3}} \exp
  \bigg[ - \frac{(\bm{R}_{k} - \bm{R}_{k - 1})^{2}}{2} 
   - \frac{(\bm{R}_{k - 1} - \bar{\bm{A}}_{k - 1})^{2}}{2 \bar{\phi}_{\infty}} \\
 & \qquad - \frac{(\bm{R}_{k} - \bar{\bm{A}}^{\dagger}_{k})^{2}}{2 \bar{\phi}_{\infty}}
   + \frac{(\bar{\bm{A}}_{k} - \bar{\bm{A}}^{\dagger}_{k})^{2}}{1 + 2 \bar{\phi}_{\infty}}
\bigg] .
\end{split}
\end{equation}
Thus we have eqs~\eqref{effective_slip_spring_size_single_subchain}-\eqref{probability_distribution_slip_spring_single_subchain} in the main text.

To calculate the statistical averages, we need the probability distribution
for the anchoring points $\lbrace \bm{A}_{k} \rbrace$. Unfortunately, the anchoring point distribution function itself
becomes rather complicated and difficult to handle. However, since
eq~\eqref{equilibrium_distribution_full_chain_equidistant_slip_spring}
is a Gaussian distribution, some statistical averages can be directly
calculated without explicitly calculating the probability distribution function.
We consider some statistical properties of the vectors which connect two
neighboring anchoring points, $\bm{U}_{k} \equiv \bm{A}_{k} - \bm{A}_{k - 1}$. This would
be interpreted as the bond vector for the anchoring points.
The probability distribution for two neighboring anchoring points can be calculated
straightforwardly:
\begin{equation}
 \begin{split}
  P_{\text{eq},\text{ES}}(\bm{A}_{k - 1},\bm{A}_{k}) 
  & \propto \int d\bm{R}_{k - 1} d\bm{R}_{k} \,
  \exp\left[ - \frac{(\bm{R}_{k} - \bm{R}_{k - 1})^{2}}{2} 
  - \frac{ (\bm{R}_{k - 1} - \bm{A}_{k - 1})^{2}}{2 \phi}
  - \frac{(\bm{R}_{k} - \bm{A}_{k})^{2}}{2 \phi} 
  \right] \\
  & \propto  \exp\left[ - \frac{\bm{U}_{k}^{2}}{2 (1 + 2 \phi)} 
  \right] .
 \end{split}
\end{equation}
Thus we have
 $\langle \bm{U}_{k} \rangle_{\text{eq}} = 0$, and
 $\langle \bm{U}_{k} \bm{U}_{k} \rangle_{\text{eq}} = (1 + 2 \phi) \bm{1}$. Two bond vectors $\bm{U}_{k}$ and $\bm{U}_{j}$
are not statistically correlated if they do not share at least one anchoring point.
Thus we have $\langle \bm{U}_{k} \bm{U}_{j} \rangle_{\text{eq}} = 0$ if $|j - k| > 1$.
If two bond vectors share one anchoring point ($j = k \pm 1$) we need the
probability distribution for three sequential anchoring points to calculate the correlation:
\begin{equation}
 \begin{split}
  P_{\text{eq},\text{ES}}(\bm{A}_{k - 1},\bm{A}_{k},\bm{A}_{k + 1}) 
  & \propto \int d\bm{R}_{k - 1}d\bm{R}_{k}d\bm{R}_{k + 1} \,
  \exp\bigg[ - \frac{(\bm{R}_{k} - \bm{R}_{k - 1})^{2}}{2} 
   - \frac{(\bm{R}_{k + 1} - \bm{R}_{k})^{2}}{2}  \\
  & \qquad   - \frac{ (\bm{R}_{k - 1} - \bm{A}_{k - 1})^{2}}{2 \phi}
  - \frac{(\bm{R}_{k} - \bm{A}_{k})^{2}}{2 \phi} 
  - \frac{(\bm{R}_{k + 1} - \bm{A}_{k + 1})^{2}}{2 \phi} 
  \bigg] \\
  & \propto  \exp\left[ - \frac{1}{2}
  \begin{bmatrix}
   \bm{U}_{k} & \bm{U}_{k + 1}
  \end{bmatrix} \cdot \bm{C}^{-1} \cdot
  \begin{bmatrix}
   \bm{U}_{k} \\ \bm{U}_{k + 1}
  \end{bmatrix} 
  \right] ,
 \end{split}
\end{equation}
where $\bm{C}^{-1}$ is the inverse of the covariance matrix,
\begin{equation}
 \label{covariance_matrix_equidistant_slip_spring_three_anchoring_points}
 \bm{C}^{-1} = \frac{1}{(1 + \phi)(1 + 3 \phi)}
  \begin{bmatrix}
   (1 + 2 \phi) \bm{1}  & \phi \bm{1} \\
   \phi \bm{1} & (1 + 2 \phi) \bm{1} 
  \end{bmatrix} .
\end{equation}
By inverting eq~\eqref{covariance_matrix_equidistant_slip_spring_three_anchoring_points},
the covariance matrix $\bm{C}$ becomes
\begin{equation}
 \bm{C} = 
  \begin{bmatrix}
   (1 + 2 \phi) \bm{1}  & - \phi \bm{1} \\
   -\phi \bm{1} & (1 + 2 \phi) \bm{1} 
  \end{bmatrix} ,
\end{equation}
and thus we have $\langle \bm{U}_{k} \bm{U}_{k \pm 1} \rangle_{\text{eq}} = - \phi \bm{1}$.
By combining these results, we have eq~\eqref{anchoring_point_bond_vector_mean_and_variance_equidistant_slip_spring}
in the main text.

\bibliographystyle{achemso}
\bibliography{slip_link_slip_spring_plateau_modulus}

\providecommand{\latin}[1]{#1}
\makeatletter
\providecommand{\doi}
  {\begingroup\let\do\@makeother\dospecials
  \catcode`\{=1 \catcode`\}=2 \doi@aux}
\providecommand{\doi@aux}[1]{\endgroup\texttt{#1}}
\makeatother
\providecommand*\mcitethebibliography{\thebibliography}
\csname @ifundefined\endcsname{endmcitethebibliography}
  {\let\endmcitethebibliography\endthebibliography}{}
\begin{mcitethebibliography}{46}
\providecommand*\natexlab[1]{#1}
\providecommand*\mciteSetBstSublistMode[1]{}
\providecommand*\mciteSetBstMaxWidthForm[2]{}
\providecommand*\mciteBstWouldAddEndPuncttrue
  {\def\EndOfBibitem{\unskip.}}
\providecommand*\mciteBstWouldAddEndPunctfalse
  {\let\EndOfBibitem\relax}
\providecommand*\mciteSetBstMidEndSepPunct[3]{}
\providecommand*\mciteSetBstSublistLabelBeginEnd[3]{}
\providecommand*\EndOfBibitem{}
\mciteSetBstSublistMode{f}
\mciteSetBstMaxWidthForm{subitem}{(\alph{mcitesubitemcount})}
\mciteSetBstSublistLabelBeginEnd
  {\mcitemaxwidthsubitemform\space}
  {\relax}
  {\relax}

\bibitem[Doi and Edwards(1986)Doi, and Edwards]{Doi-Edwards-book}
Doi,~M.; Edwards,~S.~F. \emph{The Theory of Polymer Dynamics}; Oxford
  University Press: Oxford, 1986\relax
\mciteBstWouldAddEndPuncttrue
\mciteSetBstMidEndSepPunct{\mcitedefaultmidpunct}
{\mcitedefaultendpunct}{\mcitedefaultseppunct}\relax
\EndOfBibitem
\bibitem[Fetters \latin{et~al.}()Fetters, Lohse, and
  Colby]{Fetters-Lohse-Colby-2007}
Fetters,~L.~J.; Lohse,~D.~J.; Colby,~R.~H. In \emph{Chain Dimensions and
  Entanglement Spacings}; Mark,~J.~E., Ed.; Chapter 25, pp 445--452, in {\em
  Physical Properties of Polymers Handbook}, J. E. Mark ed, 2nd ed. (Springer,
  New York, 2007)\relax
\mciteBstWouldAddEndPuncttrue
\mciteSetBstMidEndSepPunct{\mcitedefaultmidpunct}
{\mcitedefaultendpunct}{\mcitedefaultseppunct}\relax
\EndOfBibitem
\bibitem[Liu \latin{et~al.}(2006)Liu, He, van Ruymbeke, Keunings, and
  Bailly]{Liu-He-vanRuymbeke-Keunings-Bailly-2006}
Liu,~C.; He,~J.; van Ruymbeke,~E.; Keunings,~R.; Bailly,~C. Evaluation of
  different methods for the determination of the plateau modulus and the
  entanglement molecular weight. \emph{Polymer} \textbf{2006}, \emph{47},
  4461--4479\relax
\mciteBstWouldAddEndPuncttrue
\mciteSetBstMidEndSepPunct{\mcitedefaultmidpunct}
{\mcitedefaultendpunct}{\mcitedefaultseppunct}\relax
\EndOfBibitem
\bibitem[Steenbakkers \latin{et~al.}(2014)Steenbakkers, Tzoumanekas, Li, Liu,
  Kr\"{o}ger, and
  Schieber]{Steenbakkers-Tzoumanekas-Li-Liu-Kroger-Schieber-2014}
Steenbakkers,~R. J.~A.; Tzoumanekas,~C.; Li,~Y.; Liu,~W.~K.; Kr\"{o}ger,~M.;
  Schieber,~J.~D. Primitive-path statistics of entangled polymers: mapping
  multi-chain simulations onto single-chain mean-field models. \emph{New J.
  Phys.} \textbf{2014}, \emph{16}, 051027\relax
\mciteBstWouldAddEndPuncttrue
\mciteSetBstMidEndSepPunct{\mcitedefaultmidpunct}
{\mcitedefaultendpunct}{\mcitedefaultseppunct}\relax
\EndOfBibitem
\bibitem[Masubuchi \latin{et~al.}(2008)Masubuchi, Ianniruberto, Greco, and
  Marrucci]{Masubuchi-Ianniruberto-Greco-Marrucci-2008}
Masubuchi,~Y.; Ianniruberto,~G.; Greco,~F.; Marrucci,~G. Quantitative
  comparison of primitive chain network simulations with literature data of
  linear viscoelasticity for polymer melts. \emph{J. Non-Newtonian Fluid Mech.}
  \textbf{2008}, \emph{149}, 87--92\relax
\mciteBstWouldAddEndPuncttrue
\mciteSetBstMidEndSepPunct{\mcitedefaultmidpunct}
{\mcitedefaultendpunct}{\mcitedefaultseppunct}\relax
\EndOfBibitem
\bibitem[Masubuchi and Uneyama(2018)Masubuchi, and
  Uneyama]{Masubuchi-Uneyama-2018}
Masubuchi,~Y.; Uneyama,~T. Comparison among multi-chain models for entangled
  polymer dynamics. \emph{Soft Matter} \textbf{2018}, \emph{14},
  5986--5994\relax
\mciteBstWouldAddEndPuncttrue
\mciteSetBstMidEndSepPunct{\mcitedefaultmidpunct}
{\mcitedefaultendpunct}{\mcitedefaultseppunct}\relax
\EndOfBibitem
\bibitem[Everaers(1999)]{Everaers-1999}
Everaers,~R. Entanglement effects in defect-free model polymer networks.
  \emph{New J. Phys.} \textbf{1999}, \emph{1}, 12\relax
\mciteBstWouldAddEndPuncttrue
\mciteSetBstMidEndSepPunct{\mcitedefaultmidpunct}
{\mcitedefaultendpunct}{\mcitedefaultseppunct}\relax
\EndOfBibitem
\bibitem[Masubuchi \latin{et~al.}(2003)Masubuchi, Ianniruberto, Greco, and
  Marrucci]{Masubuchi-Ianniruberto-Greco-Marrucci-2003}
Masubuchi,~Y.; Ianniruberto,~G.; Greco,~F.; Marrucci,~G. Entanglement molecular
  weight and frequency response of sliplink networks. \emph{J. Chem. Phys.}
  \textbf{2003}, \emph{119}, 6925--6930\relax
\mciteBstWouldAddEndPuncttrue
\mciteSetBstMidEndSepPunct{\mcitedefaultmidpunct}
{\mcitedefaultendpunct}{\mcitedefaultseppunct}\relax
\EndOfBibitem
\bibitem[Likhtman and McLeish(2002)Likhtman, and
  McLeish]{Likhtman-McLeish-2002}
Likhtman,~A.~E.; McLeish,~T. C.~B. Quantitative Theory for Linear Dynamics of
  Linear Entangled Polymers. \emph{Macromolecules} \textbf{2002}, \emph{35},
  6332--6343\relax
\mciteBstWouldAddEndPuncttrue
\mciteSetBstMidEndSepPunct{\mcitedefaultmidpunct}
{\mcitedefaultendpunct}{\mcitedefaultseppunct}\relax
\EndOfBibitem
\bibitem[Larson \latin{et~al.}(2003)Larson, Sridhar, Leal, McKinley, Likhtman,
  and McLeish]{Larson-Sridhar-Leal-McKinley-Likhtman-McLeish-2003}
Larson,~R.~G.; Sridhar,~T.; Leal,~L.~G.; McKinley,~G.~H.; Likhtman,~A.~E.;
  McLeish,~T. C.~B. Definitions of entanglement spacing and time constants in
  the tube model. \emph{J. Rheol.} \textbf{2003}, \emph{47}, 809--818\relax
\mciteBstWouldAddEndPuncttrue
\mciteSetBstMidEndSepPunct{\mcitedefaultmidpunct}
{\mcitedefaultendpunct}{\mcitedefaultseppunct}\relax
\EndOfBibitem
\bibitem[Hua and Schieber(1998)Hua, and Schieber]{Hua-Schieber-1998}
Hua,~C.~C.; Schieber,~J.~D. Segment connectivity, chain-length breathing,
  segmental stretch, and constraint release in reptation models. I. Theory and
  single-step strain predictions. \emph{J. Chem. Phys.} \textbf{1998},
  \emph{109}, 10018--10027\relax
\mciteBstWouldAddEndPuncttrue
\mciteSetBstMidEndSepPunct{\mcitedefaultmidpunct}
{\mcitedefaultendpunct}{\mcitedefaultseppunct}\relax
\EndOfBibitem
\bibitem[Masubuchi \latin{et~al.}(2001)Masubuchi, Takimoto, Koyama,
  Ianniruberto, Greco, and
  Marrucci]{Masubuchi-Takimoto-Koyama-Ianniruberto-Greco-Marrucci-2001}
Masubuchi,~Y.; Takimoto,~J.; Koyama,~K.; Ianniruberto,~G.; Greco,~F.;
  Marrucci,~G. Brownian simulations of a network of reptating primitive chains.
  \emph{J. Chem. Phys.} \textbf{2001}, \emph{115}, 4387--4394\relax
\mciteBstWouldAddEndPuncttrue
\mciteSetBstMidEndSepPunct{\mcitedefaultmidpunct}
{\mcitedefaultendpunct}{\mcitedefaultseppunct}\relax
\EndOfBibitem
\bibitem[Doi and Takimoto(2003)Doi, and Takimoto]{Doi-Takimoto-2003}
Doi,~M.; Takimoto,~J. Molecular modelling of entanglement. \emph{Phil. Trans.
  R. Soc. Lond. A} \textbf{2003}, \emph{361}, 641--650\relax
\mciteBstWouldAddEndPuncttrue
\mciteSetBstMidEndSepPunct{\mcitedefaultmidpunct}
{\mcitedefaultendpunct}{\mcitedefaultseppunct}\relax
\EndOfBibitem
\bibitem[Likhtman(2005)]{Likhtman-2005}
Likhtman,~A.~E. Single-Chain Slip-Link Model of Entangled Polymers:
  Simultaneous Description of Neutron Spin-Echo, Rheology, and Diffusion.
  \emph{Macromolecules} \textbf{2005}, \emph{38}, 6128--6139\relax
\mciteBstWouldAddEndPuncttrue
\mciteSetBstMidEndSepPunct{\mcitedefaultmidpunct}
{\mcitedefaultendpunct}{\mcitedefaultseppunct}\relax
\EndOfBibitem
\bibitem[Nair and Schieber(2006)Nair, and Schieber]{Nair-Schieber-2006}
Nair,~D.~M.; Schieber,~J.~D. Linear Viscoelastic Predictuions of a Consistently
  Unconstrainted Brownian Slip-Link Model. \emph{Macromolecules} \textbf{2006},
  \emph{39}, 3386--3397\relax
\mciteBstWouldAddEndPuncttrue
\mciteSetBstMidEndSepPunct{\mcitedefaultmidpunct}
{\mcitedefaultendpunct}{\mcitedefaultseppunct}\relax
\EndOfBibitem
\bibitem[Chappa \latin{et~al.}(2012)Chappa, Morse, Zippelius, and
  M\"{u}ller]{Chappa-Morse-Zippelius-Muller-2012}
Chappa,~V.~C.; Morse,~D.~C.; Zippelius,~A.; M\"{u}ller,~M. Translationally
  Invariant Slip-Spring Model for Entangled Polymer Dynamics. \emph{Phys. Rev.
  Lett.} \textbf{2012}, \emph{109}, 148302\relax
\mciteBstWouldAddEndPuncttrue
\mciteSetBstMidEndSepPunct{\mcitedefaultmidpunct}
{\mcitedefaultendpunct}{\mcitedefaultseppunct}\relax
\EndOfBibitem
\bibitem[Uneyama and Masubuchi(2012)Uneyama, and
  Masubuchi]{Uneyama-Masubuchi-2012}
Uneyama,~T.; Masubuchi,~Y. Multi-chain slip-spring model for entangled polymer
  dynamics. \emph{J. Chem. Phys.} \textbf{2012}, \emph{137}, 154902\relax
\mciteBstWouldAddEndPuncttrue
\mciteSetBstMidEndSepPunct{\mcitedefaultmidpunct}
{\mcitedefaultendpunct}{\mcitedefaultseppunct}\relax
\EndOfBibitem
\bibitem[Shanbhag(2019)]{Shanbhag-2019}
Shanbhag,~S. Fast Slip Link Model for Bidisperse Linear Polymer Melts.
  \emph{Macromolecules} \textbf{2019}, \emph{52}, 3092--3103\relax
\mciteBstWouldAddEndPuncttrue
\mciteSetBstMidEndSepPunct{\mcitedefaultmidpunct}
{\mcitedefaultendpunct}{\mcitedefaultseppunct}\relax
\EndOfBibitem
\bibitem[Shanbhag(2019)]{Shanbhag-2019a}
Shanbhag,~S. Mathematical foundations of an ultra coarse-grained slip link
  model. \emph{J. Chem. Phys.} \textbf{2019}, \emph{151}, 044903\relax
\mciteBstWouldAddEndPuncttrue
\mciteSetBstMidEndSepPunct{\mcitedefaultmidpunct}
{\mcitedefaultendpunct}{\mcitedefaultseppunct}\relax
\EndOfBibitem
\bibitem[Uneyama and Masubuchi(2011)Uneyama, and
  Masubuchi]{Uneyama-Masubuchi-2011}
Uneyama,~T.; Masubuchi,~Y. Detailed Balance Condition and Effective Free Energy
  in the Primitive Chain Network Model. \emph{J. Chem. Phys.} \textbf{2011},
  \emph{135}, 184904\relax
\mciteBstWouldAddEndPuncttrue
\mciteSetBstMidEndSepPunct{\mcitedefaultmidpunct}
{\mcitedefaultendpunct}{\mcitedefaultseppunct}\relax
\EndOfBibitem
\bibitem[Schieber(2003)]{Schieber-2003}
Schieber,~J.~D. Fluctuations in entanglements of polymer liquids. \emph{J.
  Chem. Phys.} \textbf{2003}, \emph{118}, 5162--5166\relax
\mciteBstWouldAddEndPuncttrue
\mciteSetBstMidEndSepPunct{\mcitedefaultmidpunct}
{\mcitedefaultendpunct}{\mcitedefaultseppunct}\relax
\EndOfBibitem
\bibitem[Bateman(1955)]{Bateman-book}
Bateman,~H. \emph{Higher Transcendental Functions}; McGraw-Hill: New York,
  1955; Vol.~3\relax
\mciteBstWouldAddEndPuncttrue
\mciteSetBstMidEndSepPunct{\mcitedefaultmidpunct}
{\mcitedefaultendpunct}{\mcitedefaultseppunct}\relax
\EndOfBibitem
\bibitem[Evans and Morris(2008)Evans, and Morris]{Evans-Morris-book}
Evans,~D.~J.; Morris,~G.~P. \emph{Statistical Mechanics of Nonequilibrium
  Liquids}, 2nd ed.; Cambridge University Press: Cambridge, 2008\relax
\mciteBstWouldAddEndPuncttrue
\mciteSetBstMidEndSepPunct{\mcitedefaultmidpunct}
{\mcitedefaultendpunct}{\mcitedefaultseppunct}\relax
\EndOfBibitem
\bibitem[Ramirez \latin{et~al.}(2007)Ramirez, Sukumaran, and
  Likhtman]{Ramirez-Sukumaran-Likhtman-2007}
Ramirez,~J.; Sukumaran,~S.~K.; Likhtman,~A.~E. Significance of cross
  correlations in the stress relaxation of polymer melts. \emph{J. Chem. Phys.}
  \textbf{2007}, \emph{126}, 244904\relax
\mciteBstWouldAddEndPuncttrue
\mciteSetBstMidEndSepPunct{\mcitedefaultmidpunct}
{\mcitedefaultendpunct}{\mcitedefaultseppunct}\relax
\EndOfBibitem
\bibitem[Uneyama(2011)]{Uneyama-2011}
Uneyama,~T. Single Chain Slip-Spring Model for Fast Rheology Simulations of
  Entangled Polymers on {GPU}. \emph{Nihon Reoroji Gakkaishi (J. Soc. Rheol.
  Jpn.)} \textbf{2011}, \emph{39}, 135--152\relax
\mciteBstWouldAddEndPuncttrue
\mciteSetBstMidEndSepPunct{\mcitedefaultmidpunct}
{\mcitedefaultendpunct}{\mcitedefaultseppunct}\relax
\EndOfBibitem
\bibitem[Abramowitz and Stegun(1972)Abramowitz, and
  Stegun]{Abramowitz-Stegun-book}
Abramowitz,~M., Stegun,~I.~A., Eds. \emph{Handbook of Mathematical Functions
  with Formulas, Graphs, and Mathematical Tables}, 10th ed.; Dover: New York,
  1972\relax
\mciteBstWouldAddEndPuncttrue
\mciteSetBstMidEndSepPunct{\mcitedefaultmidpunct}
{\mcitedefaultendpunct}{\mcitedefaultseppunct}\relax
\EndOfBibitem
\bibitem[Olver \latin{et~al.}(2010)Olver, Lozier, Boisvert, and
  Clark]{NIST-handbook}
Olver,~F. W.~J., Lozier,~D.~W., Boisvert,~R.~F., Clark,~C.~W., Eds. \emph{NIST
  Handbook of Mathematical Functions}; Cambridge University Press, 2010\relax
\mciteBstWouldAddEndPuncttrue
\mciteSetBstMidEndSepPunct{\mcitedefaultmidpunct}
{\mcitedefaultendpunct}{\mcitedefaultseppunct}\relax
\EndOfBibitem
\bibitem[Rubinstein and Panyukov(1997)Rubinstein, and
  Panyukov]{Rubinstein-Panyukov-1997}
Rubinstein,~M.; Panyukov,~S. Nonaffine Deformation and Elasticity of Polymer
  Networks. \emph{Macromolecules} \textbf{1997}, \emph{30}, 8036--8044\relax
\mciteBstWouldAddEndPuncttrue
\mciteSetBstMidEndSepPunct{\mcitedefaultmidpunct}
{\mcitedefaultendpunct}{\mcitedefaultseppunct}\relax
\EndOfBibitem
\bibitem[Rubinstein and Panyukov(2002)Rubinstein, and
  Panyukov]{Rubinstein-Panyukov-2002}
Rubinstein,~M.; Panyukov,~S. Elasticity of Polymer Networks.
  \emph{Macromolecules} \textbf{2002}, \emph{35}, 6670--6686\relax
\mciteBstWouldAddEndPuncttrue
\mciteSetBstMidEndSepPunct{\mcitedefaultmidpunct}
{\mcitedefaultendpunct}{\mcitedefaultseppunct}\relax
\EndOfBibitem
\bibitem[Matsumoto and Nishimura(1998)Matsumoto, and
  Nishimura]{Matsumoto-Nishimura-1998}
Matsumoto,~M.; Nishimura,~T. Mersenne twister: a 623-dimensionally
  equidistributed uniform pseudo-random number generator. \emph{ACM Trans.
  Model. Comp. Simul.} \textbf{1998}, \emph{8}, 3--30,
  http://www.math.sci.hiroshima-u.ac.jp/\~{}m-mat/MT/emt.html\relax
\mciteBstWouldAddEndPuncttrue
\mciteSetBstMidEndSepPunct{\mcitedefaultmidpunct}
{\mcitedefaultendpunct}{\mcitedefaultseppunct}\relax
\EndOfBibitem
\bibitem[Devroye(1986)]{Devroye-book}
Devroye,~L. \emph{Non-Uniform Random Variate Generation}; Springer: New York,
  1986\relax
\mciteBstWouldAddEndPuncttrue
\mciteSetBstMidEndSepPunct{\mcitedefaultmidpunct}
{\mcitedefaultendpunct}{\mcitedefaultseppunct}\relax
\EndOfBibitem
\bibitem[Marsaglia and Tsang(2000)Marsaglia, and Tsang]{Marsaglia-Tsang-2000}
Marsaglia,~G.; Tsang,~W.~W. A Simple Method for Generating Gamma Variables.
  \emph{{ACM} Trans. Math. Software} \textbf{2000}, \emph{26}, 363--372\relax
\mciteBstWouldAddEndPuncttrue
\mciteSetBstMidEndSepPunct{\mcitedefaultmidpunct}
{\mcitedefaultendpunct}{\mcitedefaultseppunct}\relax
\EndOfBibitem
\bibitem[Everaers(2012)]{Everaers-2012}
Everaers,~R. Topological versus rheological entanglement length in
  primitive-path analysis protocols, tube models, and slip-link models.
  \emph{Phys. Rev. E} \textbf{2012}, \emph{86}, 022801\relax
\mciteBstWouldAddEndPuncttrue
\mciteSetBstMidEndSepPunct{\mcitedefaultmidpunct}
{\mcitedefaultendpunct}{\mcitedefaultseppunct}\relax
\EndOfBibitem
\bibitem[Masubuchi \latin{et~al.}()Masubuchi, Doi, and
  Uneyama]{Masubuchi-Doi-Uneyama-inpreparation}
Masubuchi,~Y.; Doi,~Y.; Uneyama,~T. Multi-chain slip-spring simulations with
  various slip-spring densities. arXiv:2011.03222\relax
\mciteBstWouldAddEndPuncttrue
\mciteSetBstMidEndSepPunct{\mcitedefaultmidpunct}
{\mcitedefaultendpunct}{\mcitedefaultseppunct}\relax
\EndOfBibitem
\bibitem[Masubuchi(2018)]{Masubuchi-2018}
Masubuchi,~Y. Multichain Slip-Spring Simulations for Branch Polymers.
  \emph{Macromolecules} \textbf{2018}, \emph{51}, 10184--10193\relax
\mciteBstWouldAddEndPuncttrue
\mciteSetBstMidEndSepPunct{\mcitedefaultmidpunct}
{\mcitedefaultendpunct}{\mcitedefaultseppunct}\relax
\EndOfBibitem
\bibitem[Masubuchi and Uneyama(2019)Masubuchi, and
  Uneyama]{Masubuchi-Uneyama-2019}
Masubuchi,~Y.; Uneyama,~T. Multi-chain slip-spring simulations for polyisoprene
  melts. \emph{Korea-Australia Rheology Journal} \textbf{2019}, \emph{31},
  241--248\relax
\mciteBstWouldAddEndPuncttrue
\mciteSetBstMidEndSepPunct{\mcitedefaultmidpunct}
{\mcitedefaultendpunct}{\mcitedefaultseppunct}\relax
\EndOfBibitem
\bibitem[Everaers \latin{et~al.}(2004)Everaers, Sukumaran, Grest, Svaneborg,
  Sivasubramanian, and
  Kremer]{Everaers-Sukumaran-Grest-Svaneborg-Sivasubramanian-Kremer-2004}
Everaers,~R.; Sukumaran,~S.~K.; Grest,~G.~S.; Svaneborg,~C.;
  Sivasubramanian,~A.; Kremer,~K. Rheology and Microscopic Topology of
  Entangled Polymeric Liquids. \emph{Science} \textbf{2004}, \emph{303},
  823\relax
\mciteBstWouldAddEndPuncttrue
\mciteSetBstMidEndSepPunct{\mcitedefaultmidpunct}
{\mcitedefaultendpunct}{\mcitedefaultseppunct}\relax
\EndOfBibitem
\bibitem[Sukumaran \latin{et~al.}(2005)Sukumaran, Grest, Kremer, and
  Everaers]{Sukumaran-Grest-Kremer-Everaers-2005}
Sukumaran,~S.~K.; Grest,~G.~S.; Kremer,~K.; Everaers,~R. Identifying the
  Primitive Path Mesh in Entangled Polymer Liquids. \emph{J. Polym. Sci. B:
  Polym. Phys.} \textbf{2005}, \emph{43}, 917--933\relax
\mciteBstWouldAddEndPuncttrue
\mciteSetBstMidEndSepPunct{\mcitedefaultmidpunct}
{\mcitedefaultendpunct}{\mcitedefaultseppunct}\relax
\EndOfBibitem
\bibitem[Kr\"{o}ger(2005)]{Kroger-2005}
Kr\"{o}ger,~M. Shortest multiple disconnected path for the analysis of
  entanglements in two- and three-dimensional polymeric systems. \emph{Comp.
  Phys. Comm.} \textbf{2005}, \emph{168}, 209--232\relax
\mciteBstWouldAddEndPuncttrue
\mciteSetBstMidEndSepPunct{\mcitedefaultmidpunct}
{\mcitedefaultendpunct}{\mcitedefaultseppunct}\relax
\EndOfBibitem
\bibitem[Tzoumanekas and Theodorou(2006)Tzoumanekas, and
  Theodorou]{Tzoumanekas-Theodorou-2006}
Tzoumanekas,~C.; Theodorou,~D.~N. Topological Analysis of Linear Polymer Melts:
  A Statistical Approach. \emph{Macromolecules} \textbf{2006}, \emph{39},
  4592--4604\relax
\mciteBstWouldAddEndPuncttrue
\mciteSetBstMidEndSepPunct{\mcitedefaultmidpunct}
{\mcitedefaultendpunct}{\mcitedefaultseppunct}\relax
\EndOfBibitem
\bibitem[Becerra \latin{et~al.}(2020)Becerra, C\'{o}rdoba, Katzarova, Andreev,
  Venerus, and
  Schieber]{Becerra-Cordoba-Katzarova-Andreev-Venerus-Schieber-2020}
Becerra,~D.; C\'{o}rdoba,~A.; Katzarova,~M.; Andreev,~M.; Venerus,~D.~C.;
  Schieber,~J.~D. Polymer rheology predictions from first principles using the
  slip-link model. \emph{J. Rheol.} \textbf{2020}, \emph{64}, 1035\relax
\mciteBstWouldAddEndPuncttrue
\mciteSetBstMidEndSepPunct{\mcitedefaultmidpunct}
{\mcitedefaultendpunct}{\mcitedefaultseppunct}\relax
\EndOfBibitem
\bibitem[Doi(1983)]{Doi-1983}
Doi,~M. Explanation for the 3.4-power law for viscosity of polymeric liquids on
  the basis of the tube model. \emph{J. Polym. Sci.: Polym. Phys. Ed.}
  \textbf{1983}, \emph{21}, 667--684\relax
\mciteBstWouldAddEndPuncttrue
\mciteSetBstMidEndSepPunct{\mcitedefaultmidpunct}
{\mcitedefaultendpunct}{\mcitedefaultseppunct}\relax
\EndOfBibitem
\bibitem[Fetters \latin{et~al.}(1999)Fetters, Lohse, and
  Graessley]{Fetters-Lohse-Graessley-1999}
Fetters,~L.~J.; Lohse,~D.~J.; Graessley,~W.~W. Chain Dimensions and
  Entanglement Spacings in Dense Macromolecular Systems. \emph{J. Polym. Sci.
  B: Polym. Phys.} \textbf{1999}, \emph{37}, 1023--1033\relax
\mciteBstWouldAddEndPuncttrue
\mciteSetBstMidEndSepPunct{\mcitedefaultmidpunct}
{\mcitedefaultendpunct}{\mcitedefaultseppunct}\relax
\EndOfBibitem
\bibitem[Osaki(2002)]{Osaki-2002}
Osaki,~K. A Note on Rheology of Polymeric Systems. \emph{Nihon Reoroji
  Gakkaishi (J. Soc. Rheol. Jpn.)} \textbf{2002}, \emph{30}, 165--172, in
  Japanese\relax
\mciteBstWouldAddEndPuncttrue
\mciteSetBstMidEndSepPunct{\mcitedefaultmidpunct}
{\mcitedefaultendpunct}{\mcitedefaultseppunct}\relax
\EndOfBibitem
\bibitem[Greco(2008)]{Greco-2008}
Greco,~F. Equilibrium statistical distributions for subchains in an entangled
  polymer melt. \emph{Eur. Phys. J. E} \textbf{2008}, \emph{25}, 175--180\relax
\mciteBstWouldAddEndPuncttrue
\mciteSetBstMidEndSepPunct{\mcitedefaultmidpunct}
{\mcitedefaultendpunct}{\mcitedefaultseppunct}\relax
\EndOfBibitem
\bibitem[Lebedev(1972)]{Lebedev-book}
Lebedev,~N.~N. \emph{Special Functions and Their Applications}; Dover: New
  York, 1972\relax
\mciteBstWouldAddEndPuncttrue
\mciteSetBstMidEndSepPunct{\mcitedefaultmidpunct}
{\mcitedefaultendpunct}{\mcitedefaultseppunct}\relax
\EndOfBibitem
\end{mcitethebibliography}

\clearpage

%

\end{document}